\documentclass[lettersize,journal]{IEEEtran}
\usepackage{amsmath,amsfonts}
\usepackage{algorithmic}
\usepackage{algorithm}
\usepackage{array}
\usepackage[caption=false,font=normalsize,labelfont=sf,textfont=sf]{subfig}
\usepackage{textcomp}
\usepackage{stfloats}
\usepackage{url}
\usepackage{verbatim}
\usepackage{graphicx}
\usepackage{cite}
\usepackage{color}

\usepackage{subfig}     
\usepackage{amsthm}
\usepackage{amssymb}
\usepackage{bbm}
\newtheorem{theo}{Theorem}
\newtheorem{defi}{Definition}

\newtheorem{prop}{Proposition}

\newtheorem{lem}{Lemma}

\begin{document}

\title{Delay-optimal Congestion-aware Routing and \\ Computation Offloading in Arbitrary Network}

\author{Jinkun Zhang,~\IEEEmembership{Member,~IEEE,} \quad Yuezhou Liu, \quad Edmund Yeh,~\IEEEmembership{Senior Member,~IEEE} 
\thanks{Jinkun Zhang is with the Department of Electrical and Electronic Engineering, Imperial College, London, UK (e-mail: jinkun.zhang@imperial.ac.uk). Yuezhou Liu and Edmund Yeh are with the Electrical and Computer Engineering Department, Northeastern University, Boston, USA (e-mail: eyeh@northeastern.edu).}
}

\markboth{Journal of \LaTeX\ Class Files,~Vol.~14, No.~8, August~2021}%
{Shell \MakeLowercase{\textit{et al.}}: A Sample Article Using IEEEtran.cls for IEEE Journals}


\maketitle

\begin{abstract}
Emerging edge computing paradigms enable heterogeneous devices to collaborate on complex computation applications.
However, for arbitrary heterogeneous edge networks, delay-optimal forwarding and computation offloading remains an open problem.
In this paper, we jointly optimize data/result routing and computation placement in arbitrary networks with heterogeneous node capabilities, and congestion-dependent nonlinear transmission and processing delay. 
Despite the non-convexity of the formulated problem, based on analyzing the KKT condition, we provide a set of sufficient optimality conditions that solve the problem globally.
To provide the insights for such global optimality, we show that the proposed non-convex problem is geodesic-convex with mild assumptions. We also show that the proposed sufficient optimality condition leads to a lower hemicontinuous solution set, providing stability against user-input perturbation.
We then extend the framework to incorporate utility-based congestion control and fairness. 
A fully distributed algorithm is developed to converge to the global optimum. 
Numerical results demonstrate significant improvements over multiple baselines algorithms.
\end{abstract}


\section{Introduction}
\label{Section:introduction}
\IEEEPARstart{R}{ecent} years have seen an explosion in the number of mobile and IoT devices. Many of the emerging mobile applications, such as VR/AR, autonomous driving, are computation-intensive and time-critical.
Mobile devices running these applications generate a huge amount of data traffic, which is predicted to reach 288EB per month in 2027 \cite{Ericsson2021report}. 
It is becoming impractical to direct all computation requests and their data to the central cloud due to limited backhaul bandwidth and high associated latency.
Edge computing has been proposed as a promising solution to provide computation resources and cloud-like services in close proximity to mobile devices.   
Well-known edge computing paradigms include mobile edge computing (MEC) and fog computing, which deploy computation resources at wireless access points and gateways, respectively. 

In edge computing, requesters offload their computation to the edge servers, where the network topology is typically hierarchical. 
Extending the idea of edge computing is a concept called collaborative edge computing (CEC), in which the network structure is more flexible.
In addition to point-to-point offloading, CEC permits multiple stakeholders (mobile devices, IoT devices, edge servers, or cloud) to collaborate with each other by sharing data, communication resources, and computation resources to accomplish computation tasks \cite{sahni2020multi}. CEC improves the utilization efficiency of resources so that computation-intensive and time-critical services can be better completed at the edge. Mobile devices equipped with computation capabilities can collaborate with each other through D2D communication \cite{sahni2017edge}. Edge servers can also collaborate with each other 
for load balancing or further with the central cloud to offload demands that they cannot accommodate \cite{zhu2017socially}. Furthermore, CEC is needed when there is no direct connection between devices and edge servers. Consider unmanned aerial vehicle (UAV) swarms or autonomous cars in rural areas, computation-intensive tasks of UAVs or cars far away from the wireless access point should be collaboratively computed or offloaded through multi-hop routing to the server with the help of other devices \cite{hong2019multi,sahni2017edge}. 

However, unlike traditional edge computing, CEC, or more generally, distributed computing over arbitrary network topologies, presents unique challenges in scalability, flexibility, and robustness:
(1) The scale of CEC systems can be substantial, with a large number of devices, routing paths, and concurrent tasks, requiring efficient algorithms for joint routing and computation decisions.
(2) Unlike the hierarchical and centralized structure of traditional edge computing, CEC and its control algorithm should support ad-hoc decentralized networks with flexible structures.
(3) The network environment (e.g., link status, request pattern) can be time-varying and highly heterogeneous. Algorithms must be robust and self-adaptive, with built-in support for congestion control and fairness.

To meet these challenges, we aim to develop a general framework for CEC that facilitates various types of collaboration among stakeholders. 
In particular, we consider a multi-hop network with arbitrary topology, where the nodes collaboratively finish multiple computation tasks. Nodes have heterogeneous computation capabilities and some are also data sources (sensors and mobile users) that generate data for computation tasks. Each task has a requester node for the computation result. 
We allow partial offloading introduced in \cite{wang2016mobile}, i.e., a task can be partitioned into multiple components and separately offloaded. 
e.g., in video compression, the original video can be chunked into blocks and compressed separately at multiple devices, and the results could then be merged.

Finishing a task requires the routing of data from possibly multiple data sources to multiple nodes for computation, and the routing of results to the destination (task requester). We aim for a joint routing (how to route the data/result) and computation offloading (where to compute) strategy that minimizes the total communication and computation costs.

\begin{figure}[!t]
\centering
\includegraphics[width=0.45\textwidth]{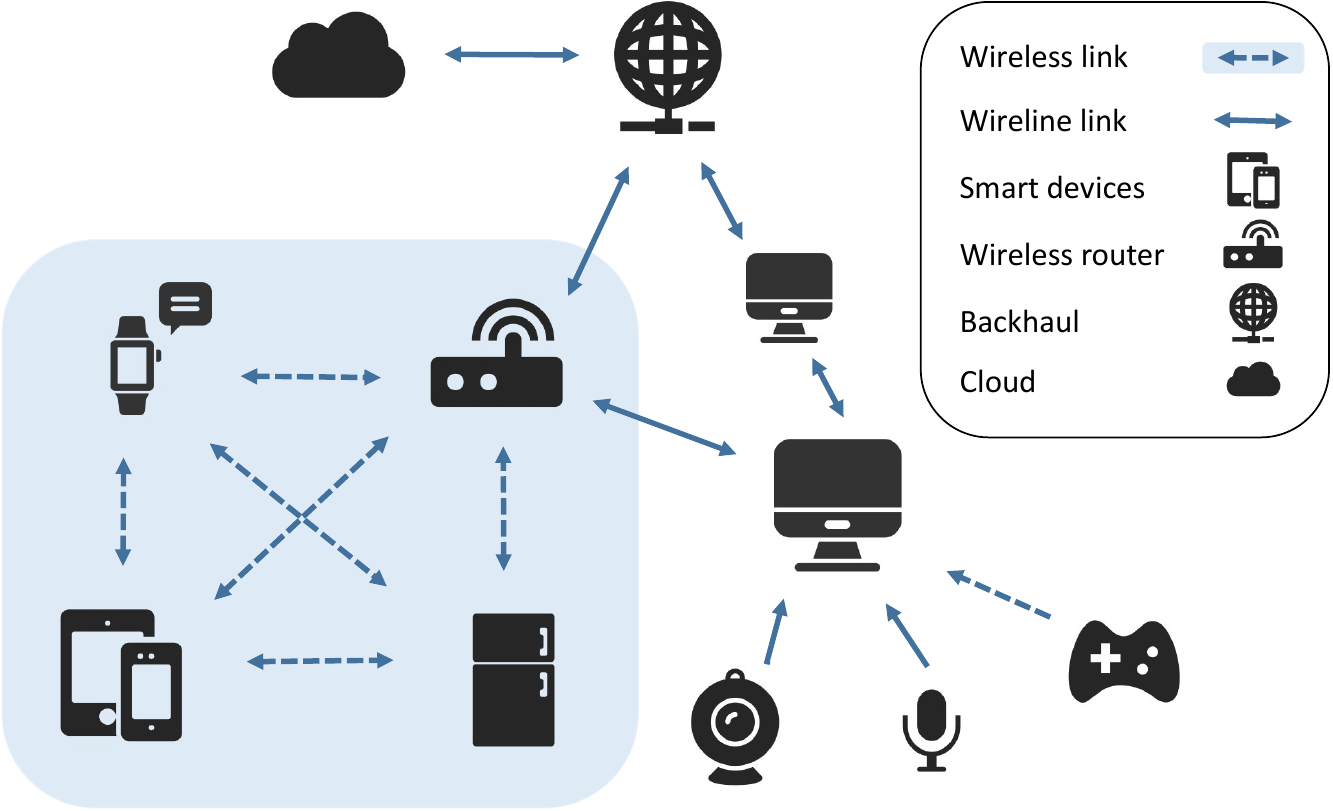}
\caption{Sample system topology involving IoT network on the edge.}
\label{fig1}
\end{figure}

Joint routing and computation offloading has been investigated by various prior contributions.  
Sahni et al. \cite{sahni2020multi} \cite{sahni2018data} adopt the model where each task is only performed once with the exact release time known, which we refer as \emph{single-instance model}. 
Zhang et al. \cite{zhang2018optimal} consider a \emph{flow model} where data collection and computation of each task are performed continuously, and time-averaged costs are measured based on the rates of data and result flows. 
Most existing studies in CEC assume the data of a tasks is provided by the requester itself \cite{hong2019multi}\cite{liu2020distributed}\cite{al2016distributed}. 
Although Sahni et al. \cite{sahni2018data}\cite{sahni2017edge} %
consider arbitrary data sources, the network is assumed to be fully connected, or with predefined routing paths.
The communication cost in previous studies is often assumed to be solely due to input data transmission \cite{luo2021qoe}\cite{sahni2020multi}, and the results are transmitted simply along the reverse path of input data 
\cite{he2021multi}\cite{funai2019computational}.
However, the size of computation results is not negligible in many practical applications,
e.g., federated or distributed machine learning~\cite{mcmahan2017communication}, file decompression, and image enhancement.
For communication, Hong et al. \cite{hong2019multi,hong2019qos} formulate heterogeneous link transmission speeds, and Sahni et al. \cite{sahni2018data} consider link bandwidth constraints. However, most works assume the link costs be linear functions of the traffic.
{Xiang et al.~\cite{xiang2020joint} study routing and computation offloading jointly with network slicing, and propose a heuristic algorithm. They consider a flow model, with non-linear delay functions, but without the consideration of computation results. }
More recently, computation and communication resource allocation has been studied through Reinforcement learning \cite{fan2023decentralized}, jointly with caching \cite{zhou2023cost}, and applied on edge learning \cite{wang2024end}, whereas congestion control and fairness are often not inherently supported.

Distinct from the above studies, in this paper, our formulation simultaneously (1) adopts the flow model on CEC network with arbitrary multi-hop topology and allows the requester node to be distinct from data sources,
2) optimizes routing for both data and results of non-negligible size,
3) models network congestion (e.g., queueing effect) by considering non-linear communication and computation costs, and
4) inherently guarantees distributed congestion control and fairness.

Specifically, we formulate a non-convex average delay minimization problem and tackle it from a distributed node-based perspective, as first introduced by \cite{gallager1977minimum}.
We first investigate the Karush–Kuhn–Tucker (KKT) necessary conditions and demonstrate that such KKT conditions can lead to arbitrarily suboptimal performance.
We then propose a set of provably sufficient conditions for global optimality by modifying the KKT condition. 
We provide novel theoretical insights to this modification by showing that our non-convex objective is geodesically convex under mild assumptions.
We also demonstrate the robustness of the proposed sufficient condition by showing it leads to a lower hemicontinuous solution set. 
We are the first to reveal such mathematical structures in this line of network optimization problems.
Based on the sufficient optimality condition, we propose a distributed and online algorithm that converges to the sufficient condition.
The algorithm is adaptive to moderate changes in network parameters.
Finally, we show that our framework can be seamlessly extended to consider distributed congestion control and fairness with global optimality intact.

Our detailed contributions are:
\begin{itemize}
    \item We formulate joint routing and computation offloading in arbitrary network with congestible links as a non-convex optimization problem. 
    \item We provide the global solution to the non-convex problem by a set of sufficient optimality conditions, and provide novel theoretical insights on such sufficiency by revealing the underlying geodesic-convexity and robustness.
    \item We seamlessly extend our global optimal solution to jointly consider congestion control and fairness. 
    \item We devise a fully distributed and adaptive algorithm, and show the advantages of the proposed algorithm through extensive experimentation, especially in congested network scenarios.
\end{itemize}

This paper is organized as follows. Section~\ref{Section:model} presents the system model and problem formulation. In Section~\ref{Section:condition}, we analyze the optimality conditions and their theoretical implications. Section~\ref{Section: Algorithm} develops a distributed and adaptive algorithm. Numerical results are presented in Section~\ref{Section: simulation}, and extensions for congestion control and fairness are discussed in Section~\ref{Section:extension}.

\section{System Model and Problem Formulation}
\label{Section:model}

We begin by presenting our formal model of a collaborative edge computing network where multiple stakeholders collaborate to carry out computation tasks. Such networks are motivated by real-word applications such as IoT networks, connected vehicles and UAV swarms.
An example that involves an IoT network at the edge is shown in Figure \ref{fig1}. 
{We summarize in Table \ref{table:tab1} the major notations used in this paper. }

\begin{table}[!t]
\caption{Major Notations}
\label{table:tab1}
\centering
\footnotesize
\begin{tabular}{|c||c|}
\hline
\textbf{Symbol} & \textbf{Definition} \\
\hline
$\mathcal{G} = (\mathcal{V},\mathcal{E})$ & Network graph with nodes $\mathcal{V}$ and links $\mathcal{E}$ \\
\hline
$\mathcal{M}$; $\mathcal{T}$ & Set of supported computation types; set of tasks \\
\hline
$(d,m)$ & A task with computation type $m$ and destination node $d$ \\
\hline
$L_{m}^{\pm}$ & Size of data and result packet for computation type $m$ \\
\hline
$r_i(d,m)$ & Exogenous input data rate for task $(d,m)$ at node $i$ \\
\hline
$t_{i}^{\pm}(d,m)$& Traffic of data  and result flows for $(d,m)$ at $i$ \\
\hline
$\phi_{ij}^{\pm}(d,m)$ & Fraction of traffic $t_{i}^{\pm}(d,m)$ forwarded to $j$ (for $j \neq 0$) \\
\hline
$\phi_{i0}^-(d,m)$ & Fraction of data traffic assigned to local CPU at $i$ \\
\hline
$f_{ij}^{\pm}(d,m)$ & Rate (packet/sec) of data and result flows on link $(i,j)$ \\
\hline
$g_i(d,m)$ & Rate (packet/sec) of data flow assigned to CPU at $i$ \\
\hline
$D_{ij}(F_{ij})$ & Transmission cost (e.g., queueing delay) on $(i,j)$ \\
\hline
$C_i(G_i)$ & Computation cost (e.g., CPU runtime) at node $i$ \\
\hline
$T(\boldsymbol{\phi})$ & Network aggregated cost\\
\hline
$\delta_{ij}^{\pm}(d,m)$ & Marginal cost for $i$ to forward data/result flow to $j$ \\
\hline
\end{tabular}
\end{table}

\subsection{Network and tasks}
We model the network with a directed and strongly connected graph $\mathcal{G} = (\mathcal{V},\mathcal{E})$, where $\mathcal{V}$ is the set of nodes 
and $\mathcal{E}$ is the set of links. 
{Nodes are assumed to be capable of both routing and computation.}
We assume that links in $\mathcal{E}$ are bidirectional, i.e., for any $(i,j) \in \mathcal{E}$, it holds that $(j,i) \in \mathcal{E}$.
Let $\mathcal{N}_i = \left\{j \big| (j,i) \in \mathcal{E}\right\} = \left\{j \big| (i,j) \in \mathcal{E}\right\}$ be the neighbors of $i$.
Computations are performed by the nodes, 
mapping input data to results of non-negligible size, e.g., image/video compression, message encoding/decoding and model training. 
Data and results for computation are transmitted through the links. 
Nodes and links are assumed to have heterogeneous computation and communication capabilities, respectively.

Communication and computation are task-driven, where a \emph{task} involves 1) forwarding input data from (potentially multiple) data sources to computation sites, 2) computing, and 3) delivering the results to a pre-specified destination. 
For example, in an IoT monitoring application, data sources could be sensors on different smart devices and the destination could be a user's cellphone. The data collected from the sensors is analyzed and processed in the network before being delivered to the user. 
We denote by $\mathcal{M}$ the set of computation types supported by the network, and a task is represented by a pair $(d,m)$,  where $d \in \mathcal{V}$ is the destination node and $m \in \mathcal{M}$ is the specified computation type. 
We denote the set of all tasks by $\mathcal{T}$ with $\mathcal{T}\subseteq\mathcal{V}\times\mathcal{M}$.

To incorporate partial offloading \cite{wang2016mobile} and measure the time-averaged network performance, for computation type $m$, we assume the exogenous input data is chunked into \emph{data packets} of equal size $L_{m}^-$. 
The data packet stream is converted into a stream of \emph{result packet} accordingly through performing computation. We assume for computation type $m$, the result packets are of equal size $L_{m}^+$.
Such assumption is adopted in many partial offloading studies that consider result size, e.g.\cite{he2021multi}, where typically $L_{m}^- \geq L_{m}^+$. 
We also allow $L_{m}^- \leq L_{m}^+$ if the result size is larger than input data, e.g., video rendering, image super-resolution or file decompression.

\subsection{Data and result flows}
In contrast to the single-instance model where each task is performed only once 
\cite{sahni2020multi}, we adopt a flow model similar to \cite{zhang2018optimal} to better capture long-term averaged network behavior.
We assume the exogenous input data packets of each task are continuously injected into the network in the form of flows with certain rates, and the computations are continuously performed.
In the network, flows of data packets, i.e. \emph{data flows}, are routed as computational input to nodes with computation resources. After computation, flows of result packets, i.e., \emph{result flows}, are generated and routed to corresponding destinations. 

We assume the exogenous input data packets of task $(d,m)$ are injected into the network with rate $r_i(d,m) \geq 0$ (packet/s) at node $i$.
Let $\boldsymbol{r} = [r_i(d,m)]_{i \in \mathcal{V}, (d,m)\in\mathcal{T}}$ be the vector of global input rates.
\footnote{Note that we allow multiple nodes $i$ for which $r_i(d,m) > 0$, representing multiple data sources; $r_d(d,m)$ can also be positive, representing computation offloading with locally provided data. } 
Let $f_{ij}^-(d,m) \geq 0$ denote the data flow rate (packet/s) on link $(i,j)$ for task $(d,m)$.
Let $g_i(d,m) \geq 0$ be the data flow rate (packet/s) forwarded to node $i$'s computation unit for task $(d,m)$, referred as the \emph{computational input rate}. 
Moreover, let $f_{ij}^+(d,m)$ be the result flow rate (packet/s) on $(i,j)$ for $(d,m)$.

\begin{figure}[!t]
\centering
\includegraphics[width=0.48\textwidth]{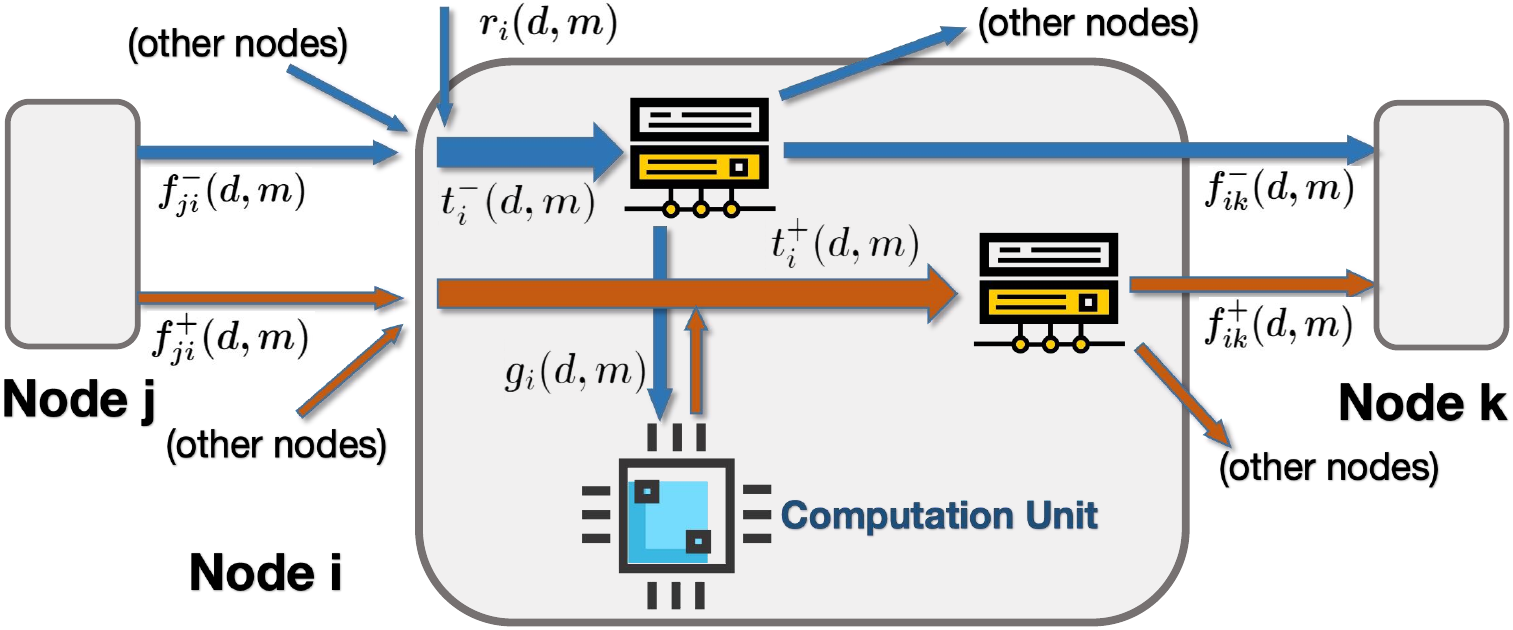}
\caption{Illustration of data and result forwarding for nodes $j \to i \to k$ for single-step computations.}
\label{fig2}
\end{figure}

We let $t_i^-(d,m)$ and $t_i^+(d,m)$  be the total data rate and total result rate for task $(d,m)$ at node $i$, respectively,
\begin{align*}
      t_i^-(d,m) &= \sum\nolimits_{j \in \mathcal{N}_i}f_{ji}^-(d,m) + r_i(d,m), 
    \\t_i^+(d,m) &= \sum\nolimits_{j \in \mathcal{N}_i}f_{ji}^+(d,m) + g_i(d,m), 
\end{align*}
Our computation flow model is illustrated in Figure~\ref{fig2}.

\subsection{Forwarding and offloading strategy}
The network performs distributed hop-by-hop multi-path packet forwarding. 
For the forwarding of data flows, we let the forwarding variable $\phi_{ij}^-(d,m) \in [0,1] $ be the fraction of data packets forwarded to node $j$ by node $i$ for task $(d,m)$.
Namely, of the traffic $t_i^-(d,m)$, we assume node $i$ forwards a fraction of $\phi_{ij}^-(d,m) \in [0,1]$ to node $j \in \mathcal{N}_i$.
Such fractional forwarding can be achieved via various methods, e.g., Random Packet Dispatching, i.e., upon receiving a data packet for $d,m$, node $i$ randomly chooses one $j \in \mathcal{N}_i$ to forward, with probability $\phi_{ij}^-(d,m)$.
Similarly, we let $\phi_{ij}^+(d,m) \in [0,1]$ be the forwarding variables of the result flow, representing the fraction of $t_i^+(d,m)$ that node $i$ forwards to $j$.
To denote computation offloading, for notation coherence, we let $\phi_{i0}^-(d,m)\in [0,1]$ be the fraction of data flow for task $(d,m)$ forwarded to the local computation unit of $i$.
Thus, 
\begin{equation*}
    \begin{aligned}
        f_{ij}^-(d,m) &= t_i^-(d,m) \phi_{ij}^-(d,m), \quad \forall j \in \mathcal{V} 
    \\ g_i(d,m) &= t_i^-(d,m) \phi_{i0}^-(d,m), 
    \\ f_{ij}^+(d,m) &= t_i^+(d,m) \phi_{ij}^+(d,m). \quad\forall j \in \mathcal{V}
    \end{aligned}
\end{equation*}
Note that $\phi_{ij}^{-}(d,m) = \phi_{ij}^{+}(d,m) \equiv 0$ if $(i,j) \not\in \mathcal{E}$. 
We denote by vector $\boldsymbol{\phi} = [\phi_{ij}^{-}(d,m), \phi_{ij}^{+}(d,m)]_{i,j \in \mathcal{V}, (d,m) \in \mathcal{T}}$ the \emph{global forwarding strategy}.

To ensure all tasks are fulfilled, every data packet must be eventually forwarded to some computation unit, and every result packet must be delivered to the corresponding destination. 
Specifically, the data flows are either forwarded to nearby nodes or to local computation unit, and the result flows exit the network at the destination.
Therefore, for all $(d,m) \in \mathcal{T}$ and $i\in \mathcal{V}$, it holds that
\begin{equation}
    \sum_{j \in \left\{0\right\} \cup \mathcal{V} } \phi_{ij}^-(d,m) = 1, \quad
    \sum_{j \in \mathcal{V} } \phi_{ij}^+(d,m) = 
    \begin{cases} 
    1, \, \text{if } i \neq d, 
    \\ 0, \, \text{if } i = d.
    \end{cases} 
    \label{FlowConservation}
\end{equation}

\subsection{Communication cost}
We denote the communication cost (e.g., average delay) on link $(i,j)$ by $D_{ij}(F_{ij})$, where $F_{ij}$ is the total flow rate (bit/s) on link $(i,j)$, given by
\begin{equation*}
    \textstyle F_{ij} = \sum\limits_{(d,m) \in \mathcal{T}} \left(L_m^-f_{ij}^-(d,m) + L_m^+f_{ij}^+(d,m)\right),
\end{equation*}
and $D_{ij}(\cdot)$ is an increasing, continuously differentiable and convex function. 
 Such convex costs subsume a variety of existing cost functions including commonly adopted linear cost\cite{ioannidis2018jointly}. It also incorporates performance metrics that reflect the network congestion status.
For example, provided that $\mu_{ij}$ is the service rate of an M/M/1 queue \cite{bertsekas2021data} with $F_{ij} < \mu_{ij}$, $D_{ij}(F_{ij}) = {F_{ij}}/\left({\mu_{ij}-F_{ij}}\right)$ gives the average number of packets waiting for or under transmission on link $(i,j)$, and is proportional to the average time for a packet from entering the queue to transmission completion.
One can also approximate the link capacity constraint $F_{ij} \leq C_{ij}$ (e.g., in \cite{liu2019joint}) by a smooth convex function that goes to infinity when approaching the capacity limit $C_{ij}$. 

\subsection{Computation cost}
We define the computation workload at node $i$ as
\vspace{-0.1\baselineskip}
\begin{equation*}
     G_i = \sum_{m\in\mathcal{M}}w_{im} \left(\sum_{d:(d,m)\in\mathcal{T}} g_i(d,m)\right),
\end{equation*}
where $w_{im} > 0$ is the weight for type $m$ at node $i$. 
We assume the computation cost at node $i$ is $C_i(G_i)$, where $C_i(\cdot)$ is an increasing, continuously differentiable and convex function.
For instance, if the computation of type $m$ requires $c_m$ CPU cycles per 
bit of input data.
By setting $w_{im} = c_m$ and $C_i(G_i) = G_i$, computation cost $C_i(G_i)$ measures the total CPU cycles. 
Alternatively, let $v_i^m$ be the computation speed of type $m$ at $i$ and $w_{im} = c_m/ v_i^m$, cost $C_i(G_i)$ measures the CPU runtime.
Function $C_i(G_i)$ can also incorporate computation congestion (e.g., average number of packets waiting for available processor or being served at CPU).
When both $D_{ij}(F_{ij})$ and $C_i(G_i)$ represent queue lengths, by Little's Law, the network aggregated cost in \eqref{aggregated_cost} is proportional to the expected packet system delay.

Note that for a network with heterogeneous computation resources, 
our model captures the fact that the workload for a certain task may be very different depending on where it is computed, e.g., some parallelizable tasks are easier at nodes employing GPU acceleration, but slower at others.

\subsection{Problem formulation}
We aim at minimizing the aggregated cost of links and devices for both communication and computation, over the forwarding and offloading strategy $\boldsymbol{\phi}$, i.e., 
  \begin{subequations}
\begin{align}
    \min_{\boldsymbol{\phi}} \quad & T(\boldsymbol{\phi}) = \sum_{(i,j) \in \mathcal{E}}D_{ij}(F_{ij}) + \sum_{i \in \mathcal{V}} C_i(G_i) 
    \label{aggregated_cost}
    \\ \text{subject to} \quad & 
    \boldsymbol{0} \leq \boldsymbol{\phi}  \leq \boldsymbol{1} \text{ and (\ref{FlowConservation}) holds.}  
\end{align}
\label{JointProblem_nodebased}
  \end{subequations}
We remark that \eqref{JointProblem_nodebased} accommodates any link or computation capacity constraints in the cost functions. 
Problem \eqref{JointProblem_nodebased} is not convex in $\boldsymbol{\phi}$.
Nevertheless, we are able to globally solve \eqref{JointProblem_nodebased} by providing a set of sufficient optimality conditions. 

\section{Optimality conditions}
\label{Section:condition}

In this section, to tackle \eqref{JointProblem_nodebased}, we first present a set of KKT necessary conditions, and then establish a set of sufficient optimality conditions based on a modification of the KKT conditions.
We provide theoretical insights to the sufficient condition by showing the geodesic convexity of \eqref{JointProblem_nodebased} under mild assumptions, and demonstrate the robustness of the proposed conditions with lower hemicontinuity.

\subsection{KKT Necessary condition}
We start by giving closed-form derivatives of $T$.
Our analysis follows \cite{gallager1977minimum} and makes non-trivial extensions to data and result flows, as well as in-network computation.

For $(i,j) \in \mathcal{E}$ and $(d,m) \in \mathcal{T}$, the marginal cost due to the increase of $\phi_{ij}^-(d,m)$ consists of two components, (1) the marginal communication cost on link $(i,j)$, and (2) the marginal cost of increasing exogenous input rate $r_j(d,m)$.
Formally, for $j \neq 0$,
\begin{subequations}
\begin{equation}
    \frac{\partial T}{\partial \phi_{ij}^-(d,m)} =  t_i^-(d,m) \left(L_{m}^{-}D_{ij}^\prime(F_{ij}) + \frac{\partial T}{\partial r_j(d,m)}\right).
\end{equation}
Similarly, the marginal cost of increasing $\phi_{i0}^-(d,m)$ consists of the marginal computation cost at $i$, and the marginal cost of increasing result traffic $t_i^+(d,m)$.
Formally,
\begin{equation}
     \frac{\partial T}{\partial \phi_{i0}^-(d,m)}= t_i^-(d,m) \left( w_{im} C_i^\prime(G_i) + \frac{\partial T}{\partial t_i^+(d,m)} \right).
\end{equation}
\label{partial_D_phi-}
\end{subequations}

Following similar reasoning, the marginal cost of increasing $\phi_{ij}^+(d,m)$ is decomposed by 
\begin{equation}
     \frac{\partial T}{\partial \phi_{ij}^+(d,m)} = 
     t_i^+(d,m) \left(L_{m}^{+}D_{ij}^\prime(F_{ij}) + \frac{\partial T}{\partial t_j^+(d,m)} \right).\label{partial_D_phi+}
\end{equation}
In the above, term $\partial T/\partial r_i(d,m)$ can be recursively expressed as a weighted sum of marginal costs for out-going links and local computation unit, namely,
\begin{equation}
    \begin{aligned}
         \frac{\partial T}{ \partial r_i(d,m)} &= \sum_{j \in \mathcal{N}_i} \phi_{ij}^-(d,m) \left(L_{m}^{-}D_{ij}^\prime(F_{ij}) + \frac{\partial T}{ \partial r_j(d,m)}\right)
    \\ &+ \phi_{i0}^-(d,m) \left(w_{im}C_i^\prime(G_i) + \frac{\partial T}{\partial t_i^+(d,m)}\right).
    \end{aligned}
    \label{partial_D_r}
\end{equation}
Similarly, the term $\partial T/\partial t_i^+(d,m)$ is given by
\begin{align}
    \frac{\partial T}{\partial t_i^+(d,m)} = \sum_{j \in \mathcal{N}_i} \phi_{ij}^+(d,m) \left(L_m^+D_{ij}^\prime(F_{ij}) +  \frac{\partial T}{\partial t_j^+(d,m)}\right).
    \label{partial_D_t}
\end{align}
One could calculate $\partial T/\partial r_i(d,m)$ and $\partial T/\partial t_i^+(d,m)$ recursively by \eqref{partial_D_r} and \eqref{partial_D_t}, since the result flow does not introduce any marginal cost at the destination, i.e., $\partial T/\partial t_d^+(d,m) = 0$.
With the presence of computation offloading, the rigorous proof of \eqref{partial_D_phi-}, \eqref{partial_D_phi+}, and \eqref{partial_D_r} is a straightforward extension of \cite{gallager1977minimum} Theorem 2, in which a pure routing problem is considered.

A set of KKT necessary conditions for the optimal solution to \eqref{JointProblem_nodebased} is given in Lemma \ref{Lemma_Necessary}.

\begin{lem}
\label{Lemma_Necessary}
Let $\boldsymbol{\phi}$ be a global optimal solution to \eqref{JointProblem_nodebased}, then for all $i \in \mathcal{V}$, $(d,m) \in \mathcal{T}$ and $j\in\left\{0\right\} \cup \mathcal{N}_i$,
\begin{equation*}
    \frac{\partial T}{ \partial \phi_{ij}^-(d,m)}  
    \begin{cases}
     = \min\limits_{k \in \left\{0\right\} \cup \mathcal{N}_i } \frac{\partial T}{ \partial \phi_{ik}^-(d,m)} , \quad \text{if } \phi_{ij}^-(d,m) >0,
    \\ \geq \min\limits_{k \in \left\{0\right\} \cup \mathcal{N}_i } \frac{\partial T}{ \partial \phi_{ik}^-(d,m)} ,  \quad \text{if } \phi_{ij}^-(d,m) =0,
    \end{cases}
\end{equation*}
and for all $j \in \mathcal{N}_i$,
\begin{equation*}
    \frac{\partial T}{ \partial \phi_{ij}^+(d,m)}  
     \begin{cases}
     = \min\limits_{k \in \mathcal{N}_i } \frac{\partial T}{ \partial \phi_{ik}^+(d,m)} , \quad \text{if } \phi_{ij}^+(d,m) >0,
    \\ \geq \min\limits_{k \in \mathcal{N}_i } \frac{\partial T}{ \partial \phi_{ik}^+(d,m)}, \quad \text{if } \phi_{ij}^+(d,m) =0.
    \end{cases}
\end{equation*}
\end{lem}
\begin{proof}
    See Appendix \ref{Proof:Lem_Necessary}.
\end{proof}

\begin{figure}[!t]
\centering
\includegraphics[width=0.48\textwidth]{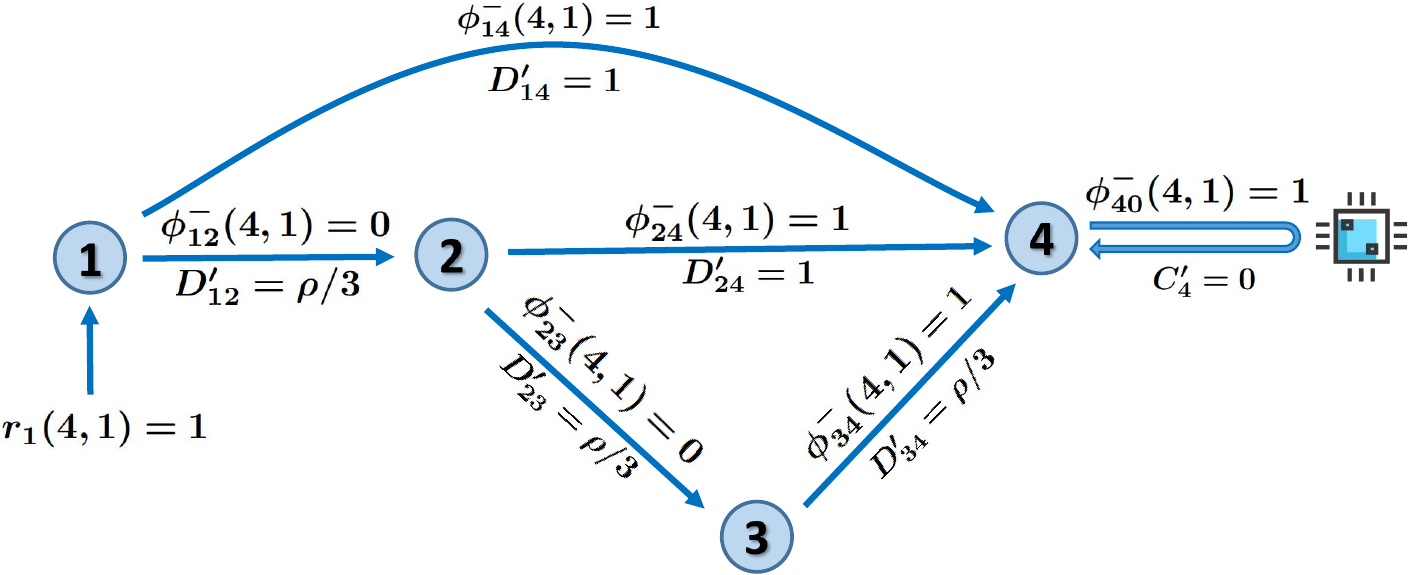}
\caption{A simple example that KKT conditions in Lemma \ref{Lemma_Necessary} leads to arbitrarily suboptimal solutions.}
\label{fig_kkt_suboptimal}
\end{figure}

The KKT conditions given in Lemma \ref{Lemma_Necessary} are not sufficient for global optimality \cite{gallager1977minimum}. 
As a matter of fact, forwarding strategies $\boldsymbol{\phi}$ satisfying the KKT conditions may perform arbitrarily worse compared to the global optimal solution.
\begin{prop}
\label{prop_arbitrarily_worse}
   For any $0<\rho<1$, there exists a scenario (i.e., network $\mathcal{G}$, tasks $\mathcal{T}$, cost functions $F_{ij}(\cdot)$, $G_i(\cdot)$, and input rates $\boldsymbol{r}$) such that $\frac{T(\boldsymbol{\phi}^*)}{T(\boldsymbol{\phi})} = \rho$, where $\boldsymbol{\phi}$ is feasible to \eqref{JointProblem_nodebased} and satisfies the condition in Lemma \ref{Lemma_Necessary}, and $\boldsymbol{\phi}^*$ is an optimal solution to \eqref{JointProblem_nodebased}.
\end{prop}
\begin{proof}
To see this, we next construct a scenario satisfying that $T(\boldsymbol{\phi}^*)/T(\boldsymbol{\phi}) = \rho$ for an arbitrarily given $ 0<\rho<1$.
Consider the simple example in Fig. \ref{fig_kkt_suboptimal}, where only one task $(d,m)$ exists with $(d,m) = (4,1)$. 
Exogenous input data occurs only at node $1$, and the computation unit is only equipped at node $4$.
For simplicity, we assume all cost functions are linear with their marginals shown in the figure. We focus solely on the communication cost by letting $C^\prime_4 = 0$.
Consider the strategy $\boldsymbol{\phi}$ shown in the figure, where the data flow is routed directly from node $1$ to node $4$, and no traffic goes through node $2$.
It can be easily verified that the condition in Lemma \ref{Lemma_Necessary} holds for the given $\boldsymbol{\phi}$, with the total cost $T(\boldsymbol{\phi}) = 1$.
However, consider another forwarding strategy $\boldsymbol{\phi}^*$: the entire data traffic is routed along path $1 \to 2 \to 3 \to 4$, incurring a total cost of $\rho$.
It is evident that $\boldsymbol{\phi}^*$ reaches the global optimum for the given network scenario, implying $T(\boldsymbol{\phi^*})/T(\boldsymbol{\phi}) = \rho$.
The ratio of $T(\boldsymbol{\phi^*})$ and $T(\boldsymbol{\phi})$ can be arbitrarily small as $\rho$ varies, that is, the KKT condition yields arbitrarily suboptimal solutions.
\end{proof}

The underlying intuition is that the KKT condition in Lemma \ref{Lemma_Necessary} automatically holds for the $i$ and $(d,m)$ with $t_i^-(d,m) = 0$ or $t_i^+(d,m) = 0$ (i.e., when no data or result traffic of task $(d,m)$ is going through node $i$), regardless of the actual forwarding strategy at node $i$. 

\subsection{Sufficient condition}
We now address the non-sufficiency of the KKT condition.
Inspired by \cite{gallager1977minimum}, we introduce a modification to the KKT condition that leads to a sufficient condition for optimality in \eqref{JointProblem_nodebased}.
Observing that for any $i$ and $(d,m)$, the traffic terms $t_i^-(d,m)$ and $t_i^+(d,m)$ repeatedly appears in RHS of \eqref{partial_D_phi-} and \eqref{partial_D_phi+} respectively for all $j \in \{0\} \cup \mathcal{V}$. 
Therefore, following \cite{gallager1977minimum}, we remove the traffic terms in the conditions given by Lemma~\ref{Lemma_Necessary}.
Such modification provably leads to a set of sufficient optimality conditions that globally solves \eqref{JointProblem_nodebased}.

\begin{theo} \label{Thm_Sufficient}
Let $\boldsymbol{\phi}$ be feasible for (\ref{JointProblem_nodebased}).  
If for all $i \in \mathcal{V}$, $(d,m) \in \mathcal{T}$ and $j\in\left\{0\right\} \cup \mathcal{N}_i$, it holds
\begin{subequations}
\label{Condition_sufficient}
\begin{equation}
    \delta_{ij}^-(d,m) \begin{cases}
    = \min\limits_{k \in \left\{0\right\}  \cup \mathcal{N}_i} \delta_{ik}^-(d,m), \quad \text{if } \phi_{ij}^-(d,m) >0,
    \\ \geq \min\limits_{k \in \left\{0\right\}  \cup \mathcal{N}_i} \delta_{ik}^-(d,m), \quad \text{if } \phi_{ij}^-(d,m) =0,
    \end{cases}
\end{equation}
and for all $j \in \mathcal{N}_i$,
\begin{equation}
     \delta_{ij}^+(d,m) \begin{cases}
    = \min\limits_{k \in \mathcal{N}_i} \delta_{ik}^+(d,m), \quad \text{if } \phi_{ij}^+(d,m) >0,
    \\ \geq \min\limits_{k \in \mathcal{N}_i} \delta_{ik}^+(d,m),  \quad \text{if } \phi_{ij}^+(d,m) =0,
    \end{cases}
\end{equation}
\end{subequations}
then $\boldsymbol{\phi}$ is a global optimal solution to \eqref{JointProblem_nodebased},

where 
$\delta_{ij}^-(d,m)$ and $\delta_{ij}^+(d,m)$ are defined as
\begin{equation}
\begin{aligned}
    \delta_{ij}^-(d,m) &= 
    \begin{cases}
    L_m^-D_{ij}^\prime(F_{ij}) + \frac{\partial T}{\partial r_j(d,m)}, &\text{if } j \neq 0,
    \\ w_{im} C_i^\prime(G_i) + \frac{\partial T}{\partial t_i^+(d,m)} , & \text{if } j =0,
    \end{cases} 
    \\ \delta_{ij}^+(d,m) &=  L_m^+D_{ij}^\prime(F_{ij}) +  \frac{\partial T}{\partial t_j^+(d,m)}.
\end{aligned}\label{delta}
\end{equation}
\end{theo}

\begin{proof}
    See Appendix \ref{proof:thm_sufficient}.
\end{proof}

To see the difference between the KKT necessary condition given by Lemma \ref{Lemma_Necessary} and the sufficient condition given by Theorem \ref{Thm_Sufficient}, consider again the example in Fig. \ref{fig_kkt_suboptimal}.
For any $\boldsymbol{\phi}$ satisfying \eqref{Condition_sufficient}, it must hold that $\phi_{12}(4,1) = 1$, $\phi_{23}(4,1) = 1$ and $\phi_{34}(4,1) = 1$, precisely indicating the shortest path $1 \to 2 \to 3 \to 4$ as expected.
Intuitively, $\delta_{ij}^-(d,m)$ and $\delta_{ij}^+(d,m)$ represent the marginal cost for node $i$ to handle a unit-rate increment of data and result traffic through forwarding to $j$, respectively. Condition \eqref{Condition_sufficient} suggests that each node handles incremental arrival flow in the way that achieves its minimum marginal cost -- either by forwarding to neighbors, or to its local CPU.

\subsection{Robustness}

We remark that condition \eqref{Condition_sufficient} is sufficient but not necessary for the global optimality of problem \eqref{JointProblem_nodebased}, i.e., \eqref{Condition_sufficient} characterizes a subset of optimal solutions to \eqref{JointProblem_nodebased}.
To see this, suppose $\boldsymbol{\phi}$ satisfies \eqref{Condition_sufficient} with $t_i^-(d,m) = 0$ for some $i$ and $(d,m)$.
In this case, forwarding variables $[\phi_{ij}^-(d,m)]_{j \in \{0\}\cup\mathcal{V}}$ can be arbitrarily perturbed without affecting feasibility or optimality, and the modified strategy may no longer satisfy \eqref{Condition_sufficient}.

Nevertheless, we argue that compared to an arbitrary optimal solution to \eqref{prop_arbitrarily_worse}, the network operator often prefers to operate at a solution that satisfies condition \eqref{Condition_sufficient} due to a practical consideration. 
This is because the input rate $\boldsymbol{r}$ is typically estimated from real-time network measurements and may vary gradually over time. In such settings, the forwarding strategy should respond smoothly to small perturbations in $\boldsymbol{r}$ while maintaining optimality. The set of strategies satisfying \eqref{Condition_sufficient} possesses this robustness property, which we next formalize as \emph{lower hemicontinuity} (LHC).

Recall that cost functions $D_{ij}(\cdot)$ and $C_i(\cdot)$ can represents link or computation capacity constraints, we denote the region of feasible link flows and workloads by
\begin{align*}
    \mathcal{D}_{F} = \bigg\{ F_{ij}, G_i \Big| &D_{ij}(F_{ij}) < \infty, \, \forall  (i,j) \in \mathcal{E}  \\&\text{ and } C_i(G_i) < \infty,\, \forall i \in \mathcal{V}\bigg\}.
\end{align*}

Let $\boldsymbol{F}(\boldsymbol{r},\boldsymbol{\phi}) = ([F_{ij}]_{(i,j)\in\mathcal{E}},[G_i]_{i \in \mathcal{V}})$ be the link flows and computation workloads for input rates $\boldsymbol{r}$ and strategy $\boldsymbol{\phi}$.
We denote the stability region of the strategy by
\begin{align*}
    \mathcal{D}_{\boldsymbol{\phi}}(\boldsymbol{r}) = \left\{\boldsymbol{\phi}\Big| \text{ \eqref{FlowConservation} holds, and } \boldsymbol{F}(\boldsymbol{r},\boldsymbol{\phi}) \in \mathcal{D}_{F}\right\}.
\end{align*}
The region of input rate $\boldsymbol{r}$ that the network can be stabilized is thus given by
\begin{align*}
    \mathcal{D}_{\boldsymbol{r}} = \left\{ \boldsymbol{r} \geq \boldsymbol{0} \Big| \mathcal{D}_{\boldsymbol{\phi}}(\boldsymbol{r}) \neq \emptyset \right\}.
\end{align*}
We assume $\mathcal{D}_{\boldsymbol{r}}$ has non-empty interior.
For $\boldsymbol{r} \in \mathcal{D}_{\boldsymbol{r}}$, let a set-valued mapping $F_{\text{opt}}: \mathcal{D}_{\boldsymbol{r}}\rightrightarrows[0,1]^{\text{dim}(\boldsymbol{\phi})}$ be $F_{\text{opt}}(\boldsymbol{r}) = \left\{\boldsymbol{\phi}:\boldsymbol{\phi} \text{ optimally solves \eqref{JointProblem_nodebased}}\right\}$; let set-valued mapping $F_{\text{suff}}: \mathcal{D}_{\boldsymbol{r}}\rightrightarrows[0,1]^{\text{dim}(\boldsymbol{\phi})}$ be $F_{\text{suff}}(\boldsymbol{r}) = \left\{\boldsymbol{\phi}:\boldsymbol{\phi} \text{ satisfies \eqref{Condition_sufficient}}\right\}$.
Then we have $F_{\text{suff}}(\boldsymbol{r}) \subseteq F_{\text{opt}}(\boldsymbol{r})$.
Moreover, to characterize stability and robustness of set-valued mapping, \emph{lower hemicontinuity} (LHC) is a generalization of semicontinuity in set-valued mappings that is widely considered in optimal control \cite{bohm1975continuity}. 

\begin{defi}[Lower Hemicontinuity]
Let $F: X \rightrightarrows Y$ be a set-valued mapping. Then $F$ is \emph{lower hemicontinuous} (LHC) at $x \in X$ if for every sequence $x_n \to x$ and every $y \in F(x)$, there exists a sequence $y_n \in F(x_n)$ such that $y_n \to y$.
\end{defi}

The intuition of LHC is that, as the input has a continuous variation, the output set does not suddenly shrink, i.e., nearby inputs always have output elements close to those of the original input.\footnote{This is different from the KKT sensitivity analysis in \cite{bertsekas1997nonlinear} Prop. 4.3.3, which gives the dynamics of single KKT point under constraint perturbation.}

\begin{prop}
\label{prop_nonLHC}
     $F_{\text{opt}}$ is not necessarily LHC over $\mathcal{D}_{\boldsymbol{r}}$.
\end{prop}
\begin{proof}
As mentioned, consider case when some $t_i^-(d,m) = 0$ in the optimal solution (which implies $r_i(d,m) = 0$),
then $\left[\phi_{ij}^-(d,m)\right]_{j \in \{0\} \cup \mathcal{V}}$ can be chosen arbitrarily without affecting the optimality. 
However, we can construct a sequence $\{\boldsymbol{r}^n\} \to \boldsymbol{r}$ such that $r_i(d,m)^n > 0$ for all $n$.
Since $r_i(d,m)^n > 0$, it must hold that $t^{-}_i(d,m)^n > 0$ through the sequence, and thus $\left[\phi^-_{ij}(d,m)^n\right]_{j \in \{0\} \cup \mathcal{V}}$ cannot be arbitrarily chosen throughout the sequence.
Therefore, those $\boldsymbol{\phi} \in F_{\text{opt}} (\boldsymbol{r})$ containing the arbitrarily chosen $\left[\phi_{ij}^-(d,m)\right]_{j \in \{0\} \cup \mathcal{V}}$ may not be a limit point of any sequence $\boldsymbol{\phi}^n \in F_{\text{opt}}(\boldsymbol{r}^n)$, violating the LHC definition.
\end{proof}

Proposition \ref{prop_nonLHC} implies the potential instability for the global optima of \eqref{JointProblem_nodebased} regarding perturbation of $\boldsymbol{r}$. 
Whereas $F_{\text{suff}}$ specified by the proposed sufficient condition \eqref{Condition_sufficient} overcomes such potential instability.

\begin{theo}
\label{thm:stability}
$F_{\text{suff}}$ is LHC over $\mathcal{D}_{\boldsymbol{r}}$.
\end{theo}

We defer the proof of Theorem \ref{thm:stability} to Section \ref{Section: Algorithm}. 
The underlying intuition is that \eqref{Condition_sufficient} includes restrictions on forwarding variables even for nodes with zero data or result traffic.

\subsection{Perspective from geodesic convexity}
\label{sec:geodesic}
Note that although \eqref{Condition_sufficient} provides global solutions to the non-convex problem \eqref{JointProblem_nodebased}, it only contains local conditions. 
Similar sufficiency was originally discovered by Gallager \cite{gallager1977minimum} and followed by others, e.g., \cite{xi2008node}.
However, the underlying mathematical structure has not been revealed.
To address this, we provide novel theoretical insight into such sufficiency by showing that with mild additional assumptions, the objective function $T(\boldsymbol{\phi})$ is geodesically convex in $\boldsymbol{\phi}$.


Note that the KKT condition and condition \eqref{Condition_sufficient} coincide for $i$ and $(d,m)$ with strictly positive $t_i^-(d,m)$ and $t_i^+(d,m)$.
Therefore, if $\boldsymbol{t} \equiv [t_i^-(d,m), t_i^+(d,m)]_{i\in\mathcal{V},(d,m)\in\mathcal{T}} > \boldsymbol{0}$, i.e., $t_i^-(d,m) > 0$ and $t_i^+(d,m)>0$ for all $i$ and $(d,m)$, condition \eqref{Condition_sufficient} and the KKT condition in Lemma \ref{Lemma_Necessary} are equivalent, implying KKT condition is both necessary and sufficient for global optimality.

\begin{prop}
\label{prop:positive_input}
    If $\boldsymbol{\phi}$ satisfies Lemma \ref{Lemma_Necessary} with $\boldsymbol{t} > \boldsymbol{0}$, then $\boldsymbol{\phi}$ optimally solves \eqref{JointProblem_nodebased}.
\end{prop}

We thus expect a stronger mathematical structure of problem \eqref{JointProblem_nodebased} beyond the general non-convexity that has not been discussed by previous works adopting similar approaches.
The concept of \emph{geodesic convex} is a natural generalization of convexity for sets and functions to Riemannian manifolds~\cite{boumal2023introduction,zhang2016first}.
In this paper, we focus solely on the case when the Riemannian manifold is a Euclidean space.\footnote{Please see Riemannian optimization textbooks, e.g., \cite{boumal2023introduction,zhang2016first}, for detailed definitions and optimization techniques of general geodesic convex functions.}
\begin{defi}[Geodesic convexity on Euclidean space]

Let $C \subset \mathbbm{R}^n$ be a compact convex set.
Function $f: C \to \mathbbm{R}$ is geodesically convex if for any $x_1, x_2 \in C$, there exists a geodesic $\gamma_{x_1x_2}(t)$ joining $x_1$ and $x_2$ with $\gamma_{x_1x_2}(t) \in C$ for $t \in [0,1]$, and $f(\gamma_{x_1x_2}(t))$ is convex with respect to $t$.
\end{defi}

To see the geodesic convexity of $T$ in $\boldsymbol{\phi}$ when $\boldsymbol{t} > 0$, 
we first rewrite \eqref{FlowConservation} using $f_{ij}^-(d,m)$, $f_{ij}^+(d,m)$, and $g_i(d,m)$,
\begin{equation}
\begin{aligned}
    \sum_{j \in \mathcal{N}_i} f_{ij}^-(d,m) + g_i(d,m) = \sum_{j\in\mathcal{N}_i}f_{ji}^-(d,m) + r_i(d,m),
    \\ \sum_{j \in \mathcal{N}_i} f_{ij}^+(d,m) = \begin{cases}
        0, \quad \text{if } i = d,
        \\ \sum_{j\in\mathcal{N}_i}f_{ji}^+(d,m) + g_i(d,m), \quad \text{o.w.}
    \end{cases}
\end{aligned}
\label{flowconservation_f_domain}
\end{equation}

We next consider the flow-domain feasible set, i.e.,  denote by $\boldsymbol{f} = [f_{ij}^-(d,m), f_{ij}^+(d,m), g_i(d,m)]_{i,j \in \mathcal{V},(d,m) \in \mathcal{T}}$ all link and CPU packet rates, then
\begin{equation*}
\begin{aligned}
    \mathcal{D}_{\boldsymbol{f}} = \left\{\boldsymbol{f}\big| \text{\eqref{flowconservation_f_domain} holds } \right\}.
\end{aligned}
\end{equation*}

For fixed $\boldsymbol{r}$, suppose $\boldsymbol{t} > \boldsymbol{0}$, then there exists a one-to-one mapping between set $\mathcal{D}_{\boldsymbol{\phi}}$ and $\mathcal{D}_{\boldsymbol{f}}$, where we denote the mapping $\boldsymbol{\phi} \to \boldsymbol{f}$ as $\boldsymbol{f}(\boldsymbol{\phi})$, and the mapping $\boldsymbol{f} \to \boldsymbol{\phi}$ as $\boldsymbol{\phi}(\boldsymbol{f})$.
Specifically, if we omit the task notation $(d,m)$, then
\begin{equation}
    \boldsymbol{\phi}(\boldsymbol{f}): \quad \,\,\,\phi_{ij}^- = \frac{f_{ij}^-}{t_i^-},\,\phi_{i0}^- = \frac{g_{i}}{t_i^-},\,\phi_{ij}^+ = \frac{f_{ij}^+}{t_i^+},
\end{equation}
and 
\begin{equation*}
\begin{aligned}
\boldsymbol{f}(\boldsymbol{\phi}): \quad
    &f_{ij}^- = t_i^- \phi_{ij}^-,\, g_i = t_i^- \phi_{i0}^-,\, f_{ij}^+= t_i^+ \phi_{ij}^+,
    \\\text{with}\quad &t_{i}^- = \sum_{v \in \mathcal{V}} r_v\sum_{p \in \mathcal{P}^-_{vi}}\prod_{l = 1}^{|p|-1}\phi^-_{p_l p_{l+1}},
    \\&t_{i}^+ = \sum_{v \in \mathcal{V}} g_v\sum_{p \in \mathcal{P}^+_{vi}}\prod_{l = 1}^{|p|-1}\phi^+_{p_l p_{l+1}}.
\end{aligned}
\end{equation*}
where $\mathcal{P}^-_{vi}$ and $\mathcal{P}^+_{vi}$ are the set of \emph{data paths} and \emph{result paths}\footnote{A \emph{data path} refers to a node sequence $p = (p_1, p_2, \cdots, p_{|p|})$ with $(p_l ,p_{l+1})\in\mathcal{E}$ and $\phi^-_{p_l p_{l+1}}>0$. A \emph{result path} refers to that with $\phi^+_{p_l p_{l+1}}>0$.
} starting from node $v$ and ends at node $i$, respectively.

Due to the convexity of $D_{ij}(\cdot)$ and $C_i(\cdot)$, we know $T$ is convex in $\boldsymbol{f}$.
Therefore, the total cost $T$ is a geodesic convex function of $\boldsymbol{\phi}$.

\begin{prop}
    Suppose $\boldsymbol{t} > \boldsymbol{0}$, Then $T(\boldsymbol{\phi})$ is geodesically convex in $\boldsymbol{\phi}$ with the geodesic function 
    \begin{equation*}
        \gamma_{\phi_1\phi_2}(t) = \boldsymbol{\phi}\left((1-t)\boldsymbol{f}(\boldsymbol{\phi}_1) + t \boldsymbol{f}(\boldsymbol{\phi}_2)\right).
    \end{equation*}
\end{prop}

When minimizing a convex function subject to linear inequality constraints, KKT conditions are necessary and sufficient for global optimality (see Proposition 4.4.1 in \cite{bertsekas1997nonlinear}).
When extended to general convex constraints, Slater's Constraint Qualification is required, i.e., there must exist a feasible point that satisfies all inequality constraints strictly (see Proposition 4.3.9 in \cite{bertsekas1997nonlinear}).
By substituting $\phi_{i0}(d,m)$ with $1 - \sum_{j \in \mathcal{V}}\phi_{ij}(d,m)$, problem \eqref{JointProblem_nodebased} satisfies the Slater's Qualification.
Recently, the sufficiency of KKT condition given Slater's Qualification is extended to Riemannian optimization \cite{jana2020convex}.
Therefore, when $\boldsymbol{t} > \boldsymbol{0}$, the KKT condition given by Lemma \ref{Lemma_Necessary} itself is sufficient for optimality without requiring condition \eqref{Condition_sufficient}.
On the other hand, when $\boldsymbol{t} > \boldsymbol{0}$ does not hold, the geodesic convexity of \eqref{JointProblem_nodebased} may no longer hold due to the existence of reflection points at $t_i^-(d,m) = 0$ or $t_i^+(d,m) = 0$, and the modification technique in Theorem \ref{Thm_Sufficient} can be adopted to eliminate the degenerate cases at these reflection points \cite{gallager1977minimum}.


\section{Distributed and Adaptive Algorithm}
\label{Section: Algorithm}

In this section, we introduce a distributed algorithm that converges to the global optimal solution of \eqref{JointProblem_nodebased} specified by Theorem \ref{Thm_Sufficient}. 
The proposed algorithm is a variant of scaled gradient projection.
It uses carefully designed scaling matrices to attain better convergence property, and is adaptive to moderate changes of exogenous input rates.
It also allows asynchronous variable update for different nodes.
Our method is based on \cite{xi2008node}, and further distinguishes data and result flows by extending the broadcasting protocol.

\subsection{Algorithm overview}
{Existence of routing loops generates redundant flow circulation, wastes network resources and causes potential instability. 
Therefore, we consider strategy $\boldsymbol{\phi}$ with \emph{loop-free} property.
For task $(d,m)$, we say $\boldsymbol{\phi}$ has a \emph{data loop} if there exists $i$ and $j$, such that $\mathcal{P}_{ij}^-(d,m)$ and $\mathcal{P}_{ji}^-(d,m)$ are both not empty, i.e., $i$ has a data path to $j$ and vice versa.
The concepts of \emph{result loop} are defined similarly with $\mathcal{P}_{ij}^+(d,m)$ and $\mathcal{P}_{ji}^+(d,m)$ being both not empty. 
We say strategy $\boldsymbol{\phi}$ is \emph{loop-free} if it has neither a data loop nor a result loop.\footnote{We allow loops concatenated by a data path and a result path of the same task, i.e., $\mathcal{P}_{ij}^-(d,m)$ and $\mathcal{P}_{ji}^+(d,m)$ are both not empty. This occurs in scenarios where the destination is the data source.}  
}


We assume the network starts with a feasible and loop-free strategy ${\boldsymbol{\phi}}^0$, where the initial total cost is finite.
Let
\begin{equation*}
    \begin{aligned}
        \boldsymbol{\phi}^-_{i}(d,m) &\equiv \left[\phi^-_{ij}(d,m)\right]_{j \in \left\{0\right\} \cup \mathcal{V}}, 
        \\\boldsymbol{\phi}^+_{i}(d,m)&\equiv\left[\phi^+_{ij}(d,m)\right]_{j \in \mathcal{V}},
    \end{aligned}
\end{equation*}
and 
\begin{equation*}
    \begin{aligned}
        \boldsymbol{\delta}^-_{i}(d,m) &\equiv \left[\delta^-_{ij}(d,m)\right]_{j \in \left\{0\right\} \cup \mathcal{V}} ,
        \\\boldsymbol{\delta}^+_{i}(d,m) &\equiv \left[\delta^+_{ij}(d,m)\right]_{j \in \mathcal{V}}.
    \end{aligned}
\end{equation*}

At the $t$-th iteration, each node $i$ independently updates its strategy for data flow, i.e., $\boldsymbol{\phi}^-_{i}(d,m)$, as follows,
\begin{equation}
    \begin{aligned}
    &\boldsymbol{\phi}^-_i(d,m)^{t+1} = 
    \\ \Big[&\boldsymbol{\phi}^-_i(d,m)^{t} -
    \left(M^-_i(d,m)^t\right)^{-1} \boldsymbol{\delta}^-_{i}(d,m)^t \Big]^+_{M^-_i(d,m)^t, \mathcal{D}_i^-(d,m)^t}
\end{aligned}
\label{scaled_gradient_projection}
\end{equation}
where $M^-_i(d,m)^t$ is a positive semi-definite diagonal scaling matrix, $\mathcal{D}_i^-(d,m)^t$ is the feasible set of $\boldsymbol{\phi}^-_i(d,m)^{t+1}$, and operator $[\cdot]^+_{A,\mathcal{D}}$ denotes the vector projection scaled by matrix $A$ onto a convex set $\mathcal{D}$, that is,
\begin{align*}
    [\boldsymbol{v}_0]^+_{A,\mathcal{D}} = \arg\min_{\boldsymbol{v} \in \mathcal{D}} (\boldsymbol{v} - \boldsymbol{v}_0)^T A (\boldsymbol{v} - \boldsymbol{v}_0). 
\end{align*}
Formula \eqref{scaled_gradient_projection} is equivalent to solving the following QP (quadratic programming)\cite{xi2008node},
\begin{equation}
\begin{aligned}
    &\boldsymbol{\phi}^-_i(d,m)^{t+1} = \underset{\boldsymbol{v} \in \mathcal{D}_i^-(d,m)^t}{\arg\min} \boldsymbol{\delta}^-_{i}(d,m)^t \cdot (\boldsymbol{v} - \boldsymbol{\phi}^-_i(d,m)^{t}) 
    \\&\quad+ (\boldsymbol{v} - \boldsymbol{\phi}^-_i(d,m)^{t})^T M^-_i(d,m)^t (\boldsymbol{v} - \boldsymbol{\phi}^-_i(d,m)^{t}),
\end{aligned} \label{variable_update}    
\end{equation}
where the feaible set $\mathcal{D}_i^-(d,m)^t$ is given by constraints $\boldsymbol{\phi}_i^-(d,m) \geq \boldsymbol{0}$ and
\begin{equation*}
\begin{gathered}
        \sum\nolimits_{ j \in \{0\} \cup \mathcal{V}}  \phi_{ij}^-(d,m) = 1; \\ \phi_{ij}^-(d,m) = 0,  \quad \forall j \in \mathcal{B}_i^-(d,m)^t.
\end{gathered}
\end{equation*}

Here $\mathcal{B}_i^-(d,m)^t$ is the set of \emph{blocked nodes} of data flow for $(d,m)$ at $i$, to guarantee the feasibility and loop-free property.
The result strategy $\boldsymbol{\phi}^+_{i}(d,m)$ is updated in a similar manner as \eqref{variable_update} with ``-'' replaced with ``+''. 
Note that $\phi_{dj}^+(d,m) \equiv 0$ for all $j \in \mathcal{V}$ 
as destinations are sinks of result flows.

The proposed method is summarized in Algorithm \ref{alg:sgp}. 
We emphasize that our method is not purely gradient-based, as the gradients are replaced by $\boldsymbol{\delta}^-_{i}(d,m)$ and $\boldsymbol{\delta}^+_{i}(d,m)$ in \eqref{scaled_gradient_projection} (recall their definition in \eqref{delta}).
We next describe in detail the calculation of $\boldsymbol{\delta}^-_{i}(d,m)$, $\boldsymbol{\delta}^+_{i}(d,m)$, scaling matrices and blocked node sets.
We then give the asynchronous convergence result and analyze the algorithm complexity. 

\begin{algorithm}[H]
\caption{Scaled Gradient Projection (SGP)}\label{alg:sgp}
\begin{algorithmic}
\STATE {\textbf{Initialize}}:
\STATE \hspace{0.5cm}Set $t \gets 0$, start with loop-free $\boldsymbol{\phi}^0$ with $T^0 < \infty$.
\STATE \hspace{0.5cm}Every node $i$ obtains $A_{ij}(T^0)$ and $A(T^0)$.
\STATE \textbf{At the end of slot $t$:}
\STATE \hspace{0.5cm}{Perform broadcast stage 1}: Compute $\partial T/\partial t_i^+(d,m)$ and \\\hspace{0.5cm}$h_i^+(d,m)$ for all $i$, $(d,m)$.
\STATE \hspace{0.5cm}{Perfrom broadcast stage 2}: Compute $\partial T/\partial r_i(d,m)$ and \\\hspace{0.5cm}$h_i^-(d,m)$ for all $i$, $(d,m)$.
\STATE \textbf{At the end of slot $t$, each node $i$:}
\STATE \hspace{0.5cm}Compute $\delta^-_{ij}(d,m)$ and $\delta^+_{ij}(d,m)$ using \eqref{delta}.
\STATE \hspace{0.5cm}Compute $\mathcal{B}^-_{ij}(d,m)$ and $\mathcal{B}^+_{ij}(d,m)$.
\STATE \hspace{0.5cm}Compute $M^+_i(d,m)$ and $M^-_i(d,m)$ using \eqref{scaling_matrix}.
\STATE \hspace{0.5cm}Solve the local optimization problem \eqref{variable_update}.
\STATE \hspace{0.5cm}Update $\boldsymbol{\phi}_i^-(d,m)$ and $\boldsymbol{\phi}_i^+(d,m)$.
\STATE \textbf{Update} {$t \gets t+1$}
\end{algorithmic}
\end{algorithm}


\subsection{Calculation of marginals}
\label{subsection: broadcast}
Each node $i$ needs to calculate vectors $\boldsymbol{\delta}^-_{i}(d,m)$ and $\boldsymbol{\delta}^+_{i}(d,m)$ following \eqref{delta}, which requires the knowledge of $D^\prime_{ij}(F_{ij})$, $C_i^\prime(G_i)$, as well as $\partial T/ \partial r_j(d,m)$ and $\partial T/ \partial t_j^+(d,m)$. 
Suppose the closed-form of $D_{ij}(\cdot)$ and $C_i(\cdot)$ are known, nodes can directly measure $D^\prime_{ij}(F_{ij})$ and $C^\prime_{i}(G_i)$ while transmitting on link $(i,j)$ and performing local computation. 
To recursively obtain $\partial T/\partial r_i(d,m)$ and $\partial T/\partial t^+_i(d,m)$ from \eqref{partial_D_r} and \eqref{partial_D_t}, respectively, a two-stage distributed broadcast protocol is introduced:
\begin{enumerate}
    \vspace{0.3\baselineskip}
  \item
  {Broadcast for $\partial T/\partial t^+_i(d,m)$:} 
  
  Node $i$ first waits until it receives messages carrying $\partial T/\partial t^+_j(d,m)$ from all its downstream neighbor, i.e., $j\in\mathcal{N}_i$
    with $\phi^+_{ij}(d,m) > 0$. 
     Then, node $i$ calculates its own $\partial T/\partial t^+_i(d,m)$ according to (\ref{partial_D_t}) with the measured $D^\prime_{ij}(F_{ij})$ and received $\partial T/\partial t^+_j(d,m)$. Next, node $i$ broadcasts $\partial T/\partial t^+_i(d,m)$ to all its upstream neighbors, i.e., $k\in\mathcal{N}_i$ with $\phi^+_{ki}(d,m) > 0$. 
     This stage starts with the destination $d$, where $d$ broadcasts $\partial T/\partial t^+_d(d,m) = 0$ to its upstream neighbors. 

\vspace{0.3\baselineskip}
\item 
{Broadcast for $\partial T/\partial r_i(d,m)$:} 

Similar as in stage 1), the exogenous input marginal $\partial T/\partial r_i(d,m)$ is calculated from \eqref{partial_D_r} recursively through broadcasting.
Note that besides all $\partial T/\partial r_j(d,m)$ from downstream neighbors, to address the case $j = 0$, node $i$ must also obtain $\partial T/\partial t^+_i(d,m)$ before calculating $\partial T/\partial r_i(d,m)$. 
Thus, the broadcast of $\partial T/\partial r_i(d,m)$ takes place after the broadcast of $\partial T/\partial t^+_i(d,m)$. 
This stage begins with the last node of each data path, where these nodes have $\partial T/\partial r_i(d,m) = w_{im}C_i^\prime(G_i) + \partial T/\partial t_i^+(d,m)$. 
\end{enumerate}

\vspace{0.3\baselineskip}
With the loop-free property, the broadcast procedure above is guaranteed to traverse throughout the network and terminate within a finite number of steps.

\subsection{ Blocked nodes and scaling matrices}
\label{subsection: Blocked nodes and scale matrices}
To achieve feasibility and the loop-free property, following \cite{gallager1977minimum}, 
we let $\mathcal{B}_i^-(d,m)^t$ be the set of nodes to which node $i$ is forbidden to forward any data flow for task $(d,m)$ at iteration $t$, and let $\mathcal{B}_i^+(d,m)^t$ be the set to which $i$ is forbidden to forward any result flow.

The intuition behind blocked nodes is as follows:
Combining Theorem \ref{Thm_Sufficient} with \eqref{partial_D_r}, if $\boldsymbol{\phi}$ is a global optimal solution to \eqref{JointProblem_nodebased}, 
$\partial T/\partial t^+_i(d,m)$ should be monotonically decreasing along any result path toward the destination node where $\partial T/\partial t^+_d(d,m) = 0$. We thus require that node $i$ should not forward any result flow to a neighbor $j$ if either 1) $\partial T/\partial t^+_j(d,m) > \partial T/\partial t^+_i(d,m)$, or 2) it could form a result path containing some link $(p,q)$ such that $\partial T/\partial t^+_q(d,m) > \partial T/\partial t^+_p(d,m)$. 
A similar requirement is applied to $\partial T/\partial r_i(d,m)$ and data paths.

Practically, the information needed to determine blocked node sets could be piggy-backed on the broadcast messages previously described with light overhead.  
The loop-free property is maintained throughout the algorithm if such a blocking mechanism is implemented in each iteration. 

The scaling matrices $M^-_i(d,m)^t$ and $M^+_i(d,m)^t$ are introduced to improve the convergence speed \cite{xi2008node}.
It also guarantees the convergence from arbitrary feasible and loop-free initial point $\boldsymbol{\phi}^0$ with finite initial cost. 
The intuition is to provide diagonal matrices that upper bound the Hessian matrices, as the Hessians are typically difficult to compute and invert.
Specifically, $M_i^+(d,m)^t$ is given by
\begin{equation}
\begin{aligned}
    &M_i^+(d,m)^t = \frac{t_i^+(d,m)^t}{2} \times \text{diag}\{A_{ij}(T^0) + 
    \\ &\left| \mathcal{N}_i \backslash \mathcal{B}_i^+(d,m)^t\right| h_j^+(d,m)^t A(T^0) \}_{j \in \mathcal{N}_i \backslash \mathcal{B}_i^+(d,m)^t},
\end{aligned}    
\label{scaling_matrix}
\end{equation}
where $T^0$ initial total cost at $\boldsymbol{\phi}^0$, $h_j^+(d,m)^t$ is the maximum path length among all existing result paths for $(d,m)$ from $j$ to destination $d$, operator $diag$ 
forms a diagonal matrix, and
\begin{align*}
    A_{ij}(T^0) = \sup_{T<T^0}D^{\prime \prime}_{ij}(F_{ij}), \quad 
     A(T^0) = \max_{(i,j) \in \mathcal{E}} A_{ij}(T^0).
\end{align*}
The definition of $M_i^-(d,m)^t$ is a repetition of the above, except with ``+'' replaced by ``-''.
Note that $h_j^+(d,m)^t$, $h_j^-(d,m)^t$ could also be piggy-backed on the broadcast messages.

\subsection{Convergence and complexity}
The proposed algorithm allows the network to update the variables one node at a time.
Such asynchrony may be caused by practical constraints such as the broadcast delay in a large-scale network. We assume that at the $t$-th iteration, only one node $i$ updates either $\boldsymbol{\phi}_i^-(d,m)$ or $\boldsymbol{\phi}_i^+(d,m)$ for one task $(d,m)\in \mathcal{T}$.  Let
\begin{align*}
    \mathcal{T}_{\boldsymbol{\phi}^-_i(d,m)} &= \left\{t \big| \text{ node } i \text{ update its } \boldsymbol{\phi}_i^-(d,m) \text{ at iteration } t \right\},
\end{align*}
with $\mathcal{T}_{\boldsymbol{\phi}^+_i(d,m)}$ defined similarly,   Then Theorem \ref{thm:convergence} holds.
\begin{theo}
\label{thm:convergence}
Assume the network start with a feasible and loop-free initial point $\boldsymbol{\phi}^0$ and the initial total cost $T^0$ is finite.
Let $\boldsymbol{\phi}^t$ be the variable generated by Algorithm \ref{alg:sgp} at the $t$-th iteration, and $T^t$ be the corresponding total cost.
Then $T^{t+1} < T^{t}$ for all $t \geq 1$.
Moreover, if
\begin{align*}
    \lim_{t \to \infty} \left|\mathcal{T}_{\boldsymbol{\phi}^-_i(d,m)}\right| = \infty, \quad \lim_{t \to \infty} \left|\mathcal{T}_{\boldsymbol{\phi}^+_i(d,m)}\right| = \infty,
\end{align*}
then the sequence $\left\{{\boldsymbol{\phi}}^t\right\}_{t\to\infty}$ converges to a limit ${\boldsymbol{\phi}}^*$, where ${\boldsymbol{\phi}}^*$ is feasible and loop-free, and $\boldsymbol{\phi}^*$ optimally solves \eqref{JointProblem_nodebased} with condition \eqref{Condition_sufficient} holding.
\end{theo}
We refer the readers to \cite{xi2008node} Theorem 2 for the proof.
With the convergence established, we now give a rigorous proof of Theorem \ref{thm:stability} in Appendix \ref{Proof:thm_stability}.

We assume that the variables of all nodes are updated once every time slot of duration $\Delta t$, and every broadcast message described in Sec. \ref{subsection: broadcast} is sent once in every slot. 
There are $2|\mathcal{E}|$ transmissions of broadcast messages corresponding to a task in one slot, and thus totally $2|\mathcal{T}||\mathcal{E}|$ transmissions per slot, with on average $2|\mathcal{T}|/\Delta t$ per link/second and at most $2\Bar{d}|\mathcal{S}|$ for each node, where $\Bar{d}$ is the largest out-degree among all nodes.
We assume the broadcast messages are sent in an out-of-band channel.
Let $t_c$ be the maximum transmission time for a broadcast message, and
$\Bar{h}$ be the maximum hop number for all data paths and result paths. 
Then the completion time of the broadcast procedure is at most $2\Bar{h}t_c$.

The number of variables for the optimization problem in \eqref{variable_update} is at most $\left(2\Bar{d}+1\right)|\mathcal{S}|$.
Each problem is a positive semidefinite diagonal QP on a simplex, which can be solved 
with polynomial complexity.

\section{Numerical Evaluation} \label{Section: simulation}

In this section, we evaluate the scaled gradient projection algorithm, i.e., \textbf{SGP} proposed in Section \ref{Section: Algorithm} by simulation. 
We implement several baseline algorithms and compare the performance of those 
against \texttt{SGP} over different networks and parameter settings. 
We also compare with a non-scaled version of \texttt{SGP} to show the improved convergence speed by using scaling matrices.

\subsection{Simulator setting}

\begin{table}[htbp]
\caption{Simulated Network Scenarios\label{tab_scenario}}
\centering
\footnotesize
\setlength{\tabcolsep}{3pt} 
\begin{tabular}{|@{\hspace{5pt}}c@{\hspace{5pt}}||@{\hspace{5pt}}c@{\hspace{5pt}}|@{\hspace{5pt}}c@{\hspace{5pt}}|@{\hspace{5pt}}c@{\hspace{5pt}}|@{\hspace{5pt}}c@{\hspace{5pt}}|@{\hspace{5pt}}c@{\hspace{5pt}}|@{\hspace{5pt}}c@{\hspace{5pt}}|@{\hspace{5pt}}c@{\hspace{5pt}}|@{\hspace{5pt}}c@{\hspace{5pt}}|}
\hline
\textbf{Network} & \multicolumn{8}{c|}{\textbf{Parameters}} \\

\textbf{Topology} & $|\mathcal{V}|$ & $|\mathcal{E}|$ & $|\mathcal{T}|$ & $|\mathcal{R}|$ & \textbf{Link} & $\Bar{d}_{ij}$ & \textbf{Comp} & $\Bar{s}_i$ \\
\hline
Connected-ER  & 20 & 40 & 15 & 5 & Queue & 10 & Queue & 12 \\
Balanced-tree & 15 & 14 & 20 & 5 & Queue & 20 & Queue & 15 \\
Fog           & 19 & 30 & 30 & 5 & Queue & 20 & Queue & 17 \\
Abilene       & 11 & 14 & 10 & 3 & Queue & 15 & Queue & 10 \\
LHC           & 16 & 31 & 30 & 5 & Queue & 15 & Queue & 15 \\
GEANT         & 22 & 33 & 40 & 7 & Queue & 20 & Queue & 20 \\
SW            & 100 & 320 & 120 & 10 & Both & 20 & Both & 20 \\
\hline
\multicolumn{9}{|c|}{\textbf{Other Parameters:} $|\mathcal{M}| = 5$, $r_{\min} = 0.5$, $r_{\max} = 1.5$} \\
\hline
\end{tabular}
\end{table}

We summarize the simulation scenarios in Table \ref{tab_scenario}. 
We evaluate \textbf{SGP} and baselines in
the following network topologies: 
\begin{itemize}
    \item \textbf{Connected-ER} is a connectivity-guaranteed Erdős-Rényi graph, generated by creating links uniformly at random with probability $p = 0.1$ on a linear network concatenating all nodes.
    
    \item \textbf{Balanced-tree} is a complete binary tree.

    \item \textbf{Fog} is a topology for fog-computing, i.e., a balanced tree with nodes on the same layer linearly linked\cite{kamran2019deco}.
    
    \item \textbf{Abilene} is the topology of the predecessor of \emph{Internet2 Network} \cite{rossi2011caching}.

    \item \textbf{GEANT} is a pan-European data network for the research and education community \cite{rossi2011caching}.

    \item \textbf{SW} (small-world) is a ring-like graph with additional short-range and long-range edges \cite{kleinberg2000small}.
\end{itemize}

\begin{figure*}[!t]
\centering
\includegraphics[width=1\textwidth]{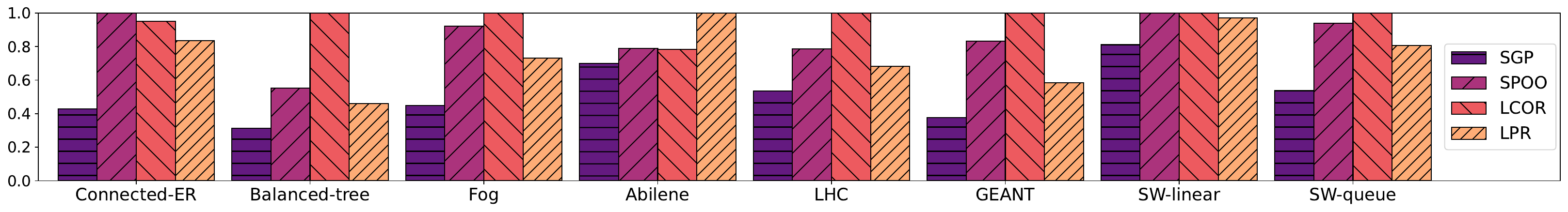}
\caption{Normalized total cost for network scenarios in Table \ref{tab_scenario}.}
\label{fig_bar}
\end{figure*}

\begin{figure*}[htbp]
  \centering
  \subfloat[\label{fig:topology}  \vspace{0\baselineskip}]{
    \includegraphics[width=0.245\textwidth]{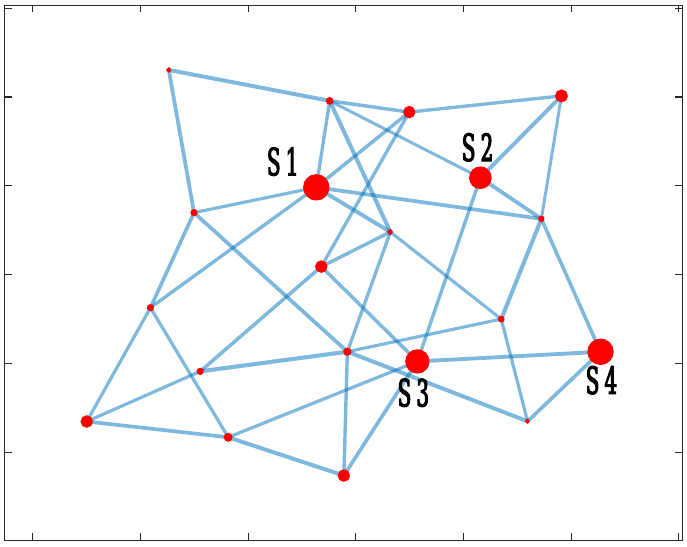}  
  }
  \subfloat[]{%
    \includegraphics[width=0.24\textwidth]{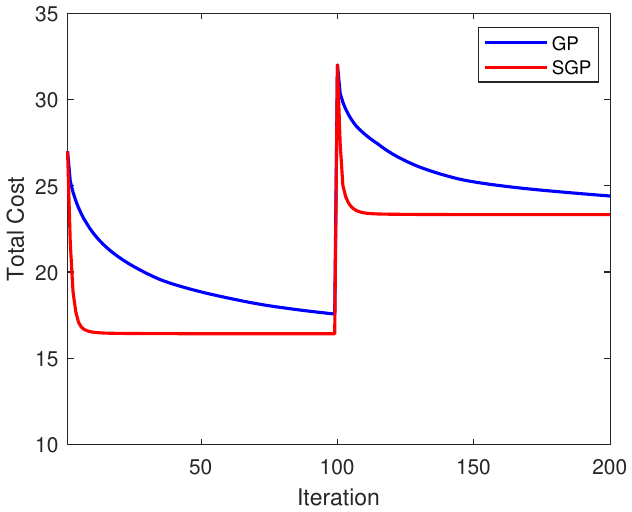}
    \label{fig:convergence}
  }
  \subfloat[]{%
    \includegraphics[width=0.24\textwidth]{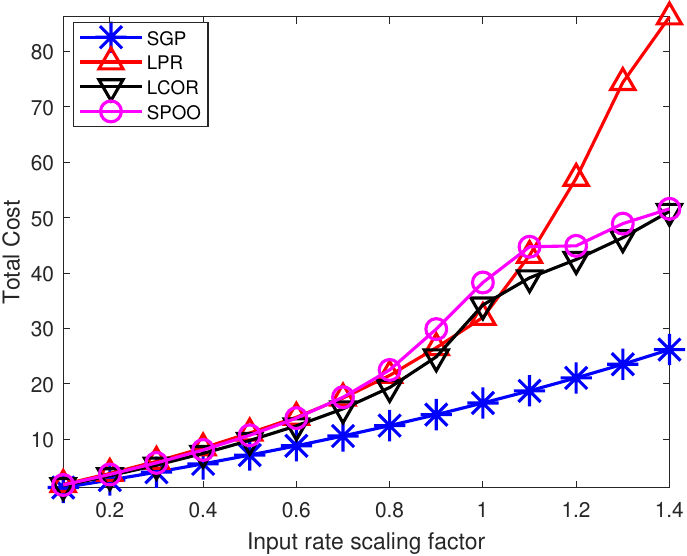}
    \label{fig:inputRate}
  }
  \subfloat[]{%
    \includegraphics[width=0.26\textwidth]{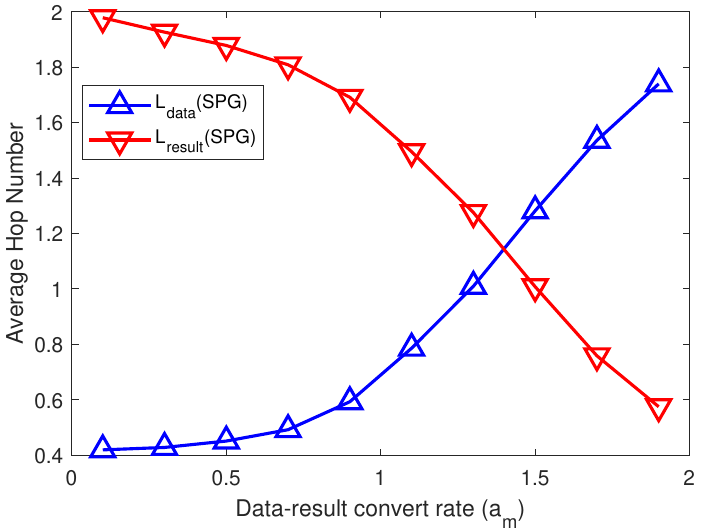}
    \label{fig:comploc}
  }
  \caption{Topology and performance results in scenario \emph{Connected-ER}.
  (a) Topology \emph{Connected-ER}, link width equal to link capacity $d_{ij}$ and node size equal to computation capacity $s_i$. (b) Convergence trajectory of \texttt{GP} and \texttt{SGP} subject to server failure at S1. (c) Total cost versus scaled input rate. (d) $L_{\text{data}}$, $L_{\text{result}}$ and their ratio versus $a_m$.}
  \label{fig:combined}
\end{figure*}

Table \ref{tab_scenario} also summarizes the number of nodes $|\mathcal{V}|$ and edges $|\mathcal{E}|$, as well as the number of 
tasks $|\mathcal{T}|$ in each network.
We assume $L^-_m = 1$ for all $m$, and let $L_m^+$ be exponentially distributed with mean value $0.5$ and truncated into the interval $[0.1,5]$, considering that most computations have result smaller than data, but special types like video rendering or generative AI models have relatively larger $L^+_m/L^-_m$.
Each task is uniformly assigned at random with one computation type and one destination node, along with $R$ random active data sources (i.e., the nodes $i$ for which $r_i(d,m) > 0$). 
The input rate $r_{i}(d,m)$ of each active data source is 
chosen u.a.r.
in $[r_{\min},r_{\max}]$.
\textbf{Link} is the type of link cost $D_{ij}(\cdot)$, where \emph{Linear} denotes a linear link cost with unit cost $d_{ij}$, i.e. $D_{ij}(F_{ij}) = d_{ij}F_{ij}$, and \emph{Queue} denotes a queueing delay with link capacity $d_{ij}$, i.e. $D_{ij}(F_{ij}) = \frac{F_{ij}}{d_{ij} - F_{ij}}$.
\textbf{Comp} is the type of computation cost $C_i(G_i)$, where \emph{Linear} denotes a weighted sum of linear cost for each type, i.e. $C_i(G_i) = s_i\sum_{m}w_{im}g_{i}^m$, and \emph{Queue} denotes a queueing delay-like computation cost with capacity $s_i$, i.e. $C_i(G_i) = \frac{\sum_{m} w_{im}g_{i}^m}{s_i - \sum_{m} w_{im}g_{i}^m}$, where the weights $w_{im}$ are 
u.a.r. drawn from $[1,5]$. 
Parameters $d_{ij}$ are 
u.a.r. drawn from $[0, 2\Bar{d}_{ij}]$. 
Parameters $s_i$ are exponential random variables with mean $\Bar{s}_i$ 
for \emph{Queue}, or uniform in $[0,2\Bar{s}_i]$ for \emph{Linear}.

We implement the following baseline algorithms: 

\begin{itemize}
    \item \textbf{GP} (Gradient Projection) is 
a non-scaled version of \texttt{SGP}, where the scaling matrices is simply chosen as follows,
\begin{small}
\begin{equation*}
\begin{aligned}
    M_i^-(d,m)^t = \frac{t_i^-(d,m)}{\beta} \times\text{diag}\left\{1,\cdots,1,0,1,\cdots,1\right\},
    \\ M_i^+(d,m)^t = \frac{t_i^+(d,m)}{\beta} \times\text{diag}\left\{1,\cdots,1,0,1,\cdots,1\right\},
\end{aligned}
\end{equation*}
\end{small}
where the only ``$0$'' entry on the diagonal of $M_i^-(d,m)^t$ is in the $j$-th position where $j = \arg\min_{k} \delta_{ik}^-(d,m)^t$, and similarly for $M_i^+(d,m)^t$.
Note that \texttt{GP} and \texttt{SGP} are both supposed to converge to global optimum of \eqref{JointProblem_nodebased}, but with different convergence speeds.

\item \textbf{SPOO} (Shortest Path Optimal Offloading) 
fixes the routing variables 
to the shortest path (measured with marginal cost at $F_{ij} = 0$, accounting for the propagation delay without queueing effect), and studies the optimal offloading along these paths. 
Namely, \texttt{SPOO} only optimize $T$ over offloading variables $\phi_{i0}^-(d,m) \in [0,1]$. It sets $\phi^-_{ij} = 1-\phi_{i0}^-$ and $\phi^+_{ij} = 1$ for $(i,j)$ on the shortest path, and sets $\phi^-_{ij} = \phi^+_{ij} = 0$ for $(i,j)$ not on the shortest path.
A similar strategy is considered in \cite{he2021multi} with linear topology and partial offloading. 

\item \textbf{LCOR} (Local Computation Optimal Routing) computes all exogenous input flows at the their data sources, 
 and optimally routes the result to destinations using scaled gradient projection in \cite{bertsekas1984second}.
That is, \texttt{LCOR} only optimizes $T$ over result routing variables $\phi_{ij}^+(d,m)$. It sets all $\phi_{i0}^-(d,m) = 1$ and $\phi_{ij}^-(d,m) = 0$. Note that we focus on the scenarios where such pure-local computation is feasible, i.e., the computation costs are finite if we set all $\phi_{i0}^-(d,m) = 1$.  

\item \textbf{LPR} (Linear Program Rounded) is the joint routing and offloading method by \cite{liu2020distributed}, which does not consider partial offloading, congestible links and result flow. 
To adapt linear link costs in  \cite{liu2020distributed} to our schemes, we use the marginal cost at zero flow. 
To ensure sufficient communication capacity for the result flow, we assign a saturation factor of $0.7$ for queueing delay costs, i.e., the data flow could not exceed $0.7$ of the real capacity.
Shortest path routing is used for result flow.
\end{itemize}

\subsection{Results and analysis}


Fig.\ref{fig_bar} compares the total cost $T$ of different algorithms in steady state over networks in Table \ref{tab_scenario} (we omit \texttt{GP} as it has the same steady state performance with \texttt{SGP}), where the bar heights of each scenario are normalized according to the worst performing algorithm. We test both linear cost and queueing delay with other parameters fixed in topology \texttt{SW}, labeled as \texttt{SW-linear} and \texttt{SW-queue}. 
Our proposed algorithm \texttt{SGP} significantly outperforms other baselines in all simulated scenarios, with as much as $50\%$ improvement over \texttt{LPR}, which also jointly optimizes routing and task offloading but does not consider partial offloading and congestible links. 
The difference of case \texttt{SW-linear} and \texttt{SW-queue} suggests that our proposed algorithm promises a considerable improvement over SOTA especially when the networks are congestible.
Note that \texttt{LCOR} and \texttt{SPOO} reflects the optimal objective for routing and offloading subproblems, respectively. The gain of jointly optimizing over both strategies could be inferred by comparing \texttt{SGP} against \texttt{LCOR} and \texttt{SPOO}. For example, \texttt{LCOR} performs very poorly in topology \emph{Balanced-tree}, because routing cannot be optimized in a tree topology. But when we switch to topology \emph{Fog} where linear links concatenate all nodes on the same depth in a balanced-tree, jointly considering routing provides much more improvement. 


We also perform refined experiments in \texttt{Connected-ER}, with the network topology and capacity shown in Fig.\ref{fig:topology}. There are $4$ major servers as labeled, and we assume server S1 fails (communication and computation capability disabled, stop performing as data source or destination) at the $100$-th iteration.
We compare the convergence speed of \texttt{GP} and \texttt{SGP} in Fig.\ref{fig:convergence} subject to S1 failure. 
\texttt{SGP} takes many fewer iterations to converge and adapt to topology change, showing the advantages of the sophisticatedly designed scaling matrices.

Fig.\ref{fig:inputRate} shows the change of total cost where all exogenous input rates $r_{i}(d,m)$ are scaled by a same factor, with other parameters fixed. The performance advantage of \texttt{SGP} quicly grows as the network becomes more congested, especially against \texttt{LPR}.

To further illustrate why \texttt{SGP} outperforms baselines significantly with congestion-dependent cost, 
we define $L_{\text{data}}$ and $L_{\text{result}}$ as the average travel distance (hop number) of data blocks from input to computation, and that of result blocks from generation to being delivered, respectively.

In Fig. \ref{fig:comploc}, we compare $L_{\text{data}}$, $L_{\text{result}}$ for \texttt{SGP} 
over different $a_m$ with other parameters fixed.
The trajectories 
suggest that the average computation offloading distance grows with $a_m \equiv L^+_m/L_m^-$. 
i.e., \texttt{SGP} tends to offload tasks generating larger result nearer to destination. 
When $a_m \gg 1$, the cost for transmitting results dominates the total cost, therefore the optimal strategy yields shorter result transmission distance.
In contrast when $a_m$ is small, the optimal strategy offloads tasks near data sources or servers, since the cost of transmitting results is low.
This phenomenon demonstrates the underlying optimality of our proposed method, reaching a ``balance'' among the cost for data forwarding, result forwarding and computation, and therefore optimizes the total cost. 

\section{Extension: Congestion Control and Fairness}
\label{Section:extension}
Thus far, our method optimally solves the joint forwarding and computation offloading problem \eqref{JointProblem_nodebased} when the exogenous input request rates $\boldsymbol{r}$ is in the stability region $\mathcal{D}_{\boldsymbol{r}}$.
There are practical situations, however, where the resulting network cost is excessive for given user demands even with the optimal forwarding and offloading strategy, since $\boldsymbol{r}$  may exceed the maximum network capacity. 
Moreover, the network operator may wish to actively balance the admitted rate for different users or tasks, in order to achieve inter-user or inter-task fairness.
Therefore, to limit and balance the exogenous input rates, we extend our proposed framework by considering an extended graph to seamlessly incorporate a utility-based congestion control.
Our congestion control method is inspired by the idea in \cite{xi2008node} to accommodate in-network computation offloading. 

\subsection{Utility-based fairness}
Extending the model in Section \ref{Section:model} where the exogenous input rates $r_i(d,m)$ are pre-defined, in this section, we assume the network operator can actively control the admitted $r_i(d,m)$ within an interval $r_i(d,m) \in [0,\Bar{r}_i(d,m)]$ for all $i \in \mathcal{V}$ and $(d,m) 0\in \mathcal{T}$, where $\Bar{r}_i(d,m) \geq 0$ is a pre-defined constant upper limit specified by network users, representing the users' maximum demand.

We associate a \emph{utility function} $U_{idm}(\cdot)$ to the exogenous inputs, where the utility of user $i$'s input for task $(d,m)$ is given by $U_{idm}(r_i(d,m))$.
We assume the utility functions $U_{idm}(\cdot)$ are monotonically increasing and concave on $[0,\Bar{r}_i(d,m)]$ with $U_{idm}(0) = 0$.
Concave $U_{idm}(\cdot)$ subsumes a variety of commonly accepted utility and fairness metrics, and is widely adopted in the literature, e.g., \cite{liu2021fair}.
For example, the $\alpha$-fairness $U(r)$ parameterized by $\alpha \geq 0$ is given by
\begin{equation*}
    U(r) = \begin{cases}
        \frac{r^{1-\alpha}}{1-\alpha}, &\quad \text{ if } 0 \leq \alpha < 1
        \\ \log (r + \epsilon), &\quad \text{ if } \alpha = 1
        \\ \frac{(r + \epsilon)^{1-\alpha}}{1-\alpha}, &\quad \text{ if } \alpha > 1
    \end{cases}
\end{equation*}
where $\epsilon$ is a positive constant.
For any $\alpha \geq 0$, the $\alpha$-fairness $U(r)$ is concave in $r$, and is strictly concave if $\alpha >0$.

Incorporating the utility metrics $U_{idm}(\cdot)$, we seek to maximize a \emph{utility-minus-cost} following \cite{kelly1997charging}, defined as
\begin{equation}
\begin{aligned}
    &\max_{\boldsymbol{r},\boldsymbol{\phi}} \quad T(\boldsymbol{r},\boldsymbol{\phi}) = \sum_{i \in \mathcal{V}}\sum_{(d,m)\in\mathcal{T}} U_{idm}(r_i(d,m)) 
    \\ & \quad \quad \quad \quad - \sum_{(i,j)\in \mathcal{E}} D_{ij}(F_{ij}) - \sum_{i \in \mathcal{V}} C_{i}(G_i) 
    \\ &\text{subject to} \quad \boldsymbol{r} \in \mathcal{D}_{\boldsymbol{r}}, \quad  \boldsymbol{\phi} \in \mathcal{D}_{\boldsymbol{\phi}}(\boldsymbol{r}).
\end{aligned}
\label{CongestionControlObj}
\end{equation}

{
We remark that by assuming individual utilities for every combination of $i$ and $(d,m)$, we consider the \emph{inter-user inter-task} fairness, where the admitted rate of each user node $i$ and each task $(d,m)$ is balanced by maximizing the aggregated utility.
Alternatively, one could consider solely the \emph{inter-user} fairness by imposing utility $U_i(\cdot)$ on the total admitted rate at $i$ given by $\sum_{(d,m) \in \mathcal{T}}r_i(d,m)$, or solely the \emph{inter-task} fairness by imposing utility $U_{dm}(\cdot)$ on the total admitted rate of task $(d,m)$ given by $\sum_{i \in \mathcal{V}} r_i(d,m)$.
In this paper, we solve \eqref{CongestionControlObj}.
}

\subsection{Extended network}
Problem \eqref{CongestionControlObj} can be optimally solved by extending our network 
 model in Section \ref{Section:model}. 
Consider an \emph{extended network} denoted by graph $\mathcal{G}^E = (\mathcal{V}^E,\mathcal{E}^{E})$,
where $\mathcal{V}^E = \mathcal{V} \cup \mathcal{V}^V$ denotes the physical nodes $\mathcal{V}$ and a set of \emph{virtual nodes} $\mathcal{V}^V$, 
and $\mathcal{E}^E =  \mathcal{E} \cup \mathcal{E}^V$ denotes the original links $\mathcal{E}$ and a set of \emph{virtual links} $\mathcal{E}^V$.

The virtual node set $\mathcal{V}^V$ consists of $|\mathcal{V}|$ nodes, each corresponding to one physical node, serving as a ``gateway'' of request admission. We denote by $v^V$ the virtual node corresponding to physical node $v$. 
Set $\mathcal{E}^V$ consists of the virtual links coming out of the virtual nodes. Specifically, we assume the exogenous input requests are migrated from physical nodes to their corresponding virtual nodes, and the input rate of task $(d,m)$ at virtual node $i^V$ is fixed to the upper limit $\Bar{r}_i(d,m)$.
Virtual node $i^V$ has a virtual out-link $(i^V, i)$ connecting to the corresponding physical node, on which the actual admitted flow of rate $r_i(d,m)$ is forwarded.
For the remaining rate $(\Bar{r}_i(d,m) - r_i(d,m))$ that is rejected by the real network, we assume it is admitted by $i^V$, directly converted to result flow, and sent to the destination $d$ through another virtual link $(i^V,d)$.
We denote by $f^V_{i}(d,m)$ the flow on the virtual link $(i^V,d)$, i.e., $f^V_i(d,m) = \Bar{r}_i(d,m) - r_i(d,m)$. 

Let $\phi_{i^V i}(d,m)$ denote the fraction of admitted rate at virtual node $i^V$ for task $(d,m)$, and let $\phi_{i^V d}(d,m)$ denote the fraction of rejected rate. 
If $\Bar{r}_i(d,m) > 0$, it holds that $\phi_{i^V i}(d,m) = r_i(d,m) / \Bar{r}_i(d,m)$ and $\phi_{i^V d}(d,m) = f^V_i(d,m) / \Bar{r}_i(d,m)$.
Then the flow conservation in \eqref{FlowConservation} is augmented with
\begin{equation}
    \phi_{i^V i}(d,m) + \phi_{i^V d}(d,m) = 1, \quad \forall (d,m) \in \mathcal{T}, i \in \mathcal{V}. 
    \label{flowconservation_phi_congestion}
\end{equation}

\begin{figure}[t!]
\centerline{
\includegraphics[width=0.45\textwidth]{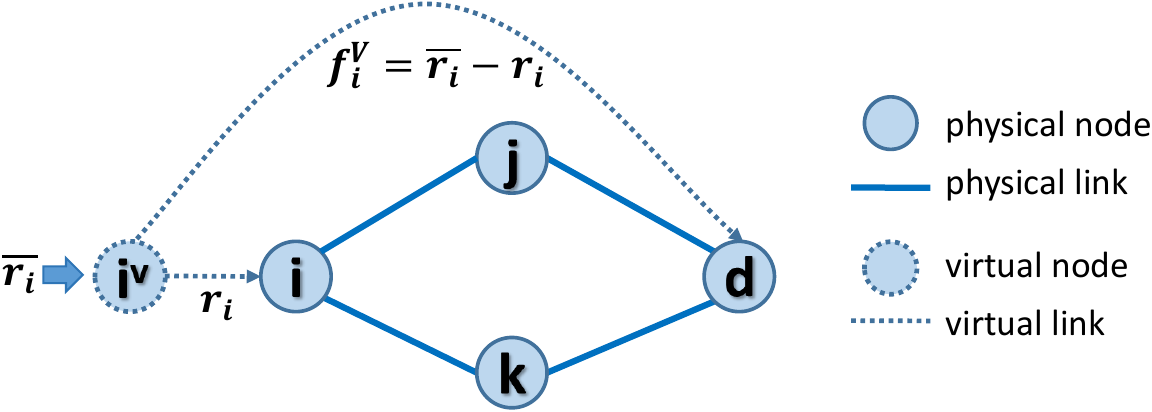}
}
\caption{Illustration of extended graph. The input rate upper limit $\Bar{r}_i$ is admitted by the virtual gate $i^{\text{V}}$. Among this, rate $r_i$ is further admitted by the physical node $i$. The rest is forwarded along the virtual link $(i^\text{V},d)$ and converted to the result stage.} 
\label{fig_congestion_control}
\end{figure}

We next assign link costs on the virtual links in $\mathcal{E}^V$.
For virtual link $(i^V,i)$, we do not assume any cost, i.e., $D_{i^Vi}(\cdot) \equiv 0$.
For virtual link $(i^V,d)$, we assume 
\begin{equation*}
\begin{aligned}
    &D_{(i^V,d)}(f^V_i(d,m)) = U_{idm}(\Bar{r}_i(d,m)) - U_{idm}(r_i(d,m))
    \\ &= U_{idm}(\Bar{r}_i(d,m)) - U_{idm}\left(\Bar{r}_i(d,m) - f^V_i(d,m)\right).
\end{aligned}
\end{equation*}
Namely, $D_{(i^V,d)}(f^V_i(d,m))$ represents the loss of utility due to rejected requests.
Due to the concavity of $U_{idm}$, we know $D_{(i^V,d)}(f^V_i(d,m))$ is increasing convex in $f^V_i(d,m)$, coherent with our previous assumption on $D_{ij}(F_{ij})$ in Section \ref{Section:model}.

Therefore, the utility-minus-cost maximization problem \eqref{CongestionControlObj} is equivalent to the following cost minimization problem on the extended graph $\mathcal{G}^E$, 
\begin{equation}
\begin{aligned}
        &\min_{\boldsymbol{\phi}^E} \quad T^E(\boldsymbol{\phi}^E) \equiv \sum_{(i,j)\in \mathcal{E}^E} D_{ij}(F_{ij}) + \sum_{i \in \mathcal{V}} C_{i}(G_i) 
    \\ &\text{subject to} \quad  \boldsymbol{\phi}^E \in \mathcal{D}_{\boldsymbol{\phi}^E},
\end{aligned}
\label{CongestionControlObj_eqv}
\end{equation}
where $\boldsymbol{\phi}^E$ includes both physical forwarding variables $\boldsymbol{\phi}$ and virtual node forwarding fractions $[\phi_{i^V i}(d,m), \phi_{i^V d}(d,m)]$. Set $\mathcal{D}_{\boldsymbol{\phi}^E}$ is defined by \eqref{FlowConservation} and \eqref{flowconservation_phi_congestion}.

\subsection{Optimal admitted rates}
Although problem \eqref{CongestionControlObj_eqv} is also non-convex, observing that  \eqref{CongestionControlObj_eqv} shares identical mathematical form with \eqref{JointProblem_nodebased}, we can extend condition \eqref{Condition_sufficient} to globally solve \eqref{CongestionControlObj_eqv}.

\begin{theo}[Sufficient Condition with Congestion Control]
\label{Thm_Sufficient_congestion}
Let $\boldsymbol{\phi}^E \in \mathcal{D}_{\boldsymbol{\phi}^E}$.  
If condition \eqref{Condition_sufficient} holds, and the following holds for all $i^V \in \mathcal{V}^V$ and $(d,m) \in \mathcal{T}$, 
\begin{equation}
\begin{aligned}
    &\delta_{i^V i}(d,m) \leq \delta_{i^V d}(d,m), \quad \text{if } \phi_{i^V i}(d,m) > 0,
    \\ &\delta_{i^V i}(d,m) \geq \delta_{i^V d}(d,m), \quad \text{if } \phi_{i^V d}(d,m) > 0,
\end{aligned}
\label{Condition_sufficient_congestion}
\end{equation}
then $\boldsymbol{\phi}^E$ is a global optimal solution to \eqref{CongestionControlObj_eqv},
where
\begin{equation*}
\begin{aligned}
    \delta_{i^V i}(d,m) = \frac{\partial T}{ \partial r_i(d,m)}, \quad
     \delta_{i^V d}(d,m) = 
    U_{idm}^\prime(r_i(d,m)).
\end{aligned}
\end{equation*}
\end{theo}

The proof of Theorem \ref{Thm_Sufficient_congestion} is omitted as it is almost a repetition of Theorem \ref{Thm_Sufficient}.
The above sufficient condition \eqref{Condition_sufficient_congestion} for congestion control can be intuitively interpreted as the following:
upon receiving a newly arrived input data packet $(d,m)$ at node $i$, the congestion control gateway $i^V$ compares the marginal network cost if the packet is admitted, i.e., $\partial T/ \partial r_i(d,m)$, and the marginal utility loss if the packet is rejected, i.e., $U_{idm}^\prime(r_i(d,m))$.
The arrival packet is then admitted if the former is smaller and rejected if not. Theorem \ref{Thm_Sufficient_congestion} implies that such a local admission policy leads to a global optimal solution.

The proposed Algorithm \ref{alg:sgp} is naturally extendable to incorporate congestion control in a distributed and adaptive manner.
The implementation of each virtual node $i^V$ is carried out by the corresponding physical node $v$ with light overhead.
Remark that Algorithm \ref{alg:sgp} itself does not specify how to find a feasible initial state $\boldsymbol{\phi}^0$. 
When extended to congestion control, however, a feasible initial state $(\boldsymbol{\phi}^E)^0$ is naturally introduced by setting $\phi_{i^V i}(d,m) = 0$ for all $i$ and $a$, i.e., the extended algorithm can always start with rejecting any arrival packets.

\section{Conclusion} \label{Section:Conclusion}
We propose a novel joint routing and computation offloading model incorporating the result flow, partial offloading and multi-hop routing for both data and result. 
To the best of our knowledge, this is also the first flow model analysis of computation offloading with congestion-dependent link cost and arbitrary network topology. 
We propose a non-convex total cost minimization problem and optimally solve it by providing sufficient optimality conditions. 
We provide novel theoretical insights into the sufficient condition by introducing geodesic convexity, and demonstrate its robustness through the concept of lower hemicontinuity. 
We devise a fully distributed and scalable algorithm that reaches the global optimal.
We compare our proposed algorithm with several baseline methods, observing a significant improvement in all tested scenarios. 
Finally, our framework can be seamlessly extended to incorporate congestion control and inter-user inter-task fairness with global optimality intact.

\appendix

\subsection{Proof of Lemma \ref{Lemma_Necessary}}
\label{Proof:Lem_Necessary}
The Lagrangian function of problem \eqref{JointProblem_nodebased} is given by
\begin{equation*}
    \begin{gathered}
L(\boldsymbol{\phi},\boldsymbol{\lambda},\boldsymbol{\mu}) = T(\boldsymbol{\phi}) - \sum_{i\in\mathcal{V}}\sum_{ (d,m)\in\mathcal{T}} \Bigg[
        \\ \lambda^-_{idm} \left( \sum_{j \in \{0\} \cup \mathcal{V}} \phi_{ij}^-(d,m) -1\right)
        \\ + \lambda^+_{idm} \left( \sum_{j \in \mathcal{V}} \phi_{ij}^+(d,m) - \mathbbm{1}_{i \neq d} \right)
        \\  + \left(\sum_{j\in\{0\}\cup\mathcal{V}}\mu^-_{ijdm}\phi^-_{ij}(d,m) + \sum_{j\in\mathcal{V}}\mu^+_{ijdm}\phi^+_{ij}(d,m)\right)\Bigg],
    \end{gathered}
\end{equation*}
where $\boldsymbol{\lambda} = [\lambda^-_{idm}, \lambda^+_{idm}]_{i,d,m}$ and $\boldsymbol{\mu} = [\mu^-_{ijdm},\mu^+_{ijdm}]_{i,j,d,m}$ with $\lambda^{\pm}_{idm} \in \mathbb{R}$ and $\mu^{\pm}_{ijdm} \geq 0$ are the Lagrangian multipliers corresponding to constraint \eqref{FlowConservation} and $\boldsymbol{\phi} \geq \boldsymbol{0}$, respectively.

Suppose $\boldsymbol{\phi}$ is a global optimal solution to \eqref{JointProblem_nodebased}, then there must exist a set of $(\boldsymbol{\lambda},\boldsymbol{\mu})$ such that \cite{bertsekas1997nonlinear}
\begin{gather*}
    \frac{\partial L}{\partial \phi^-_{ij}(d,m)} = 0, \quad \frac{\partial L}{\partial \phi^+_{ij}(d,m)} = 0,
    \\\mu^-_{ijdm}\phi^-_{ij}(d,m) = 0, \quad \mu^+_{ijdm}\phi^+_{ij}(d,m) = 0.
\end{gather*}
By Section \ref{Section:condition}, for this set of $(\boldsymbol{\lambda},\boldsymbol{\mu})$, it holds that
\begin{equation*}
\begin{aligned}
   \frac{\partial L}{\partial \phi_{ij}^{\pm}(d,m)} &= \frac{\partial T}{\partial \phi^{\pm}_{ij}(d,m)} - \lambda^\pm_{idm} - \mu^{\pm}_{ijdm}.
\end{aligned}
\end{equation*}
Combining above with the complementary slackness $\mu^{\pm}_{ijdm}\phi^{\pm}_{ij}(d,m) = 0$ (i.e., when $\phi^{\pm}_{ij}(d,m) > 0$, it must hold that $\mu^{\pm}_{ijdm} = 0$), and notice the arbitrariness of $\mu^{\pm}_{ijdm}$ when $\phi^{\pm}_{ij}(d,m) = 0$, we know that 
\begin{equation*}
    \frac{\partial T}{\partial \phi^{\pm}_{ij}(d,m)} \begin{cases}
        = \lambda^{\pm}_{idm}, \quad \text{if } \phi^{\pm}_{ij}(d,m) > 0,
        \\> \lambda^{\pm}_{idm} \quad \text{if } \phi^{\pm}_{ij}(d,m) = 0.
    \end{cases}
\end{equation*}
Therefore, Lemma \ref{Lemma_Necessary} holds by taking 
\begin{equation*}
\begin{gathered}
    \lambda^{-}_{idm} = \min_{j \in \{0\}\cup\mathcal{N}_i} \partial T/\partial \phi_{ij}^-(d,m), \\ \lambda^{-}_{idm} = \min_{j \in \mathcal{N}_i} \partial T/\partial \phi_{ij}^+(d,m)
\end{gathered}
\end{equation*}
in the above.

\subsection{Proof of Theorem \ref{Thm_Sufficient}}
\label{proof:thm_sufficient}

For simplicity, we consider the non-destination nodes in this proof. Namely, we assume $\sum_{j} \phi_{ij}^+ = 1$ for all $i$, while the derivation is applicable to destination nodes by enforcing those $\phi_{ij}^+\equiv0$.
By \eqref{partial_D_r} and \eqref{delta},  we have 
\begin{align*}
    \frac{\partial T}{\partial r_i(d,m)} &= \sum_{j \in \left\{0\right\} \cup \mathcal{N}_i} \phi_{ij}^-(d,m) \delta_{ij}^-(d,m) 
    \\ &= \sum_{j : \phi_{ij} > 0} \phi_{ij}^-(d,m) \lambda_{idm}^- 
    \\ &= \lambda_{idm}^-,
\end{align*}
where $\lambda^{-}_{idm} = \min_{j \in \{0\}\cup\mathcal{N}_i} \partial T/\partial \phi_{ij}^-(d,m)$ is the Lagrangian multiplier for the constraint $\sum_{j} \phi_{ij}^- = 1$. Thus when \eqref{Condition_sufficient} holds, we have for all $i$ and $(d,m)$,
\begin{align}
    \delta_{ij}^-(d,m) \geq \frac{\partial T}{\partial r_i(d,m)}, \forall j \in \left\{0 \right\} \cup \mathcal{N}_i. \label{SuffCond_alt1}
\end{align}
Similarly we have for all $i$ and $(d,m)$,
\begin{align}
    \delta_{ij}^+(d,m) \geq \frac{\partial T}{\partial t_i^+(d,m)}, \quad \forall j \in \mathcal{N}_i. \label{SuffCond_alt2}
\end{align}

To prove $ \boldsymbol{\phi}$ satisfying \eqref{Condition_sufficient} minimizes $$T = \sum_{(i,j) \in \mathcal{E}}D_{ij}(F_{ij}) + \sum_{i \in \mathcal{V}}C_i(G_i),$$ let $\boldsymbol{\phi}^* \neq \boldsymbol{\phi}$ be another feasible set of variable, and with corresponding packet rates link flows and computation workload $F_{ij}^*, \forall (i,j) \in \mathcal{E}$ and $G_i^*, \forall i \in \mathcal{V}$.
Given both $\boldsymbol{\phi}$ and $\boldsymbol{\phi}^*$ are both feasible routing/offloading strategies, we know $(F_{ij},G_i)$ and $(F_{ij}^*,G_i^*)$ are in the feasible set (a convex polytope) of the flow model problem (\ref{JointProblem}). 
\begin{align}
    \min_{f^-,f^+,g} \quad & T = \sum_{(i,j) \in \mathcal{E}} D_{ij}(F_{ij}) + \sum_{i \in \mathcal{V}}C_i(G_i) \label{JointProblem}
    \\ \text{such that} \quad &  \text{\eqref{flowconservation_f_domain} hold,} \nonumber
    \\ & g_i(d,m) \geq 0, \, f_{ij}^-(d,m) \geq 0, \,f_{ij}^+(d,m) \geq 0, \nonumber
\end{align}
Due to the convexity of the feasible set of \eqref{JointProblem}, for any $\mu \in [0,1]$, $\left( (1 - \mu) F_{ij} + \mu F_{ij}^* ,  (1 - \mu) G_i + \mu G_i^* \right)$ is also feasible for (\ref{JointProblem}), we then let
\begin{align*}
   T(\mu) &= \sum_{(i,j) \in \mathcal{E}}D_{ij}(  (1 - \mu) F_{ij} + \mu F_{ij}^* ) 
   \\ &+ \sum_{i \in \mathcal{V}}C_i((1 - \mu) G_i + \mu G_i^*). 
\end{align*}
Since $T$ is convex in $f^-$, $f^+$ and $g$, we know $T(\mu)$ is convex in $\mu$. Thus combining with the arbitrary choice of $\boldsymbol{\phi}^*$, the sufficiency in Theorem \ref{Thm_Sufficient} is proved if $\frac{d T(\mu)}{d \mu}$ is non-negative at $\mu = 0$. That is, we will show the following is non-negative
\begin{equation}
\begin{aligned}
    \frac{d T(\mu)}{ d \mu} \bigg|_{\mu = 0} &= \sum_{(i,j) \in \mathcal{E}}D^\prime_{ij}(F_{ij}) (F_{ij}^* - F_{ij}) 
    \\ &+ \sum_{i \in \mathcal{V}} C^\prime_i(G_i) (G_i^{*} - G_i). 
\end{aligned}    
\label{proof_obj}
\end{equation}

Starting with the data flow, multiply both side of (\ref{SuffCond_alt1}) by $\phi_{ij}^{-*}(d,m)$ and sum over $j \in \left\{ 0 \right\} \cup \mathcal{N}_i $, we have
\begin{equation}
\begin{gathered}
    w_{im}C^\prime_i(G_i) \phi_{i0}^{-*}(d,m) + \sum_{j \in \mathcal{N}_i} L_m^-D_{ij}^\prime(F_{ij}) \phi_{ij}^{-*}(d,m) 
    \\ \geq \frac{\partial T}{\partial r_i(d,m)} - \frac{\partial T}{\partial t_i^+(d,m)} \phi_{i0}^{-*}(d,m) 
    \\ - \sum_{j \in \mathcal{N}_i} \frac{ \partial T}{\partial r_j(d,m)} \phi_{ij}^{-*}(d,m),
\end{gathered}    
\label{proof_ineqToEq1}
\end{equation}
then multiply both side by $t_i^{-*}(d,m) = \sum_{j \in \mathcal{N}_i}f_{ji}^{-*}(d,m) + r_i(d,m)$, we have
\begin{align*}
     &w_{im}C^\prime_i(G_i) g_i^{*}(d,m) + L_m^-\sum_{j \in \mathcal{N}_i} D_{ij}^\prime(F_{ij}) f_{ij}^{-*}(d,m)
    \\ \geq &  t_i^{-*}(d,m) \frac{\partial T}{\partial r_i(d,m)} -  \frac{\partial T}{\partial t_i  ^+(d,m)} t_i^{-*}(d,m) \phi_{i0}^{-*}(d,m) 
    \\ - &\sum_{j \in \mathcal{N}_i} \frac{ \partial T}{\partial r_j(d,m)} t_i^{-*}(d,m) \phi_{ij}^{-*}(d,m),
\end{align*}
further sum over $(d,m) \in \mathcal{T}$ and $i \in \mathcal{V}$, the RHS of above  becomes
\begin{equation*}
\begin{gathered}
    \sum_{i \in \mathcal{V}} \sum_{(d,m) \in \mathcal{T}} w_{im}C^\prime_i(G_i) g_i^{*}(d,m) 
    \\ + \sum_{(i,j) \in \mathcal{E}} \sum_{(d,m) \in \mathcal{T}} L_m^- D_{ij}^\prime(F_{ij}) f_{ij}^{-*}(d,m)
    \\= \sum_{i\in\mathcal{V}} C^\prime_i(G_i) G_i^* + \sum_{(i,j) \in \mathcal{E}}D^\prime_{ij}(F_{ij})F^{-*}_{ij}
\end{gathered}
\end{equation*}
thus
\begin{equation}
\begin{aligned}
    &\sum_{i\in\mathcal{V}} C^\prime_i(G_i) G_i^* + \sum_{(i,j) \in \mathcal{E}}D^\prime_{ij}(F_{ij})F^{-*}_{ij}
    \\ &\geq \sum_{i \in \mathcal{V}} \sum_{(d,m) \in \mathcal{T}} t_i^{-*}(d,m) \frac{\partial T}{\partial r_i(d,m)} 
    \\ &- \sum_{i \in \mathcal{V}} \sum_{(d,m) \in \mathcal{T}}    \frac{\partial T}{\partial t_i^+(d,m)} t_i^{-*}(d,m) \phi_{i0}^{-*}(d,m)
    \\  &- \sum_{(d,m) \in \mathcal{T}} \sum_{i \in \mathcal{V}} \sum_{j \in \mathcal{N}_i } \frac{ \partial T}{\partial r_j(d,m)} t_i^{-*}(d,m) \phi_{ij}^{-*}(d,m),
\end{aligned}   
\label{proof_sum_1}
\end{equation}
where $F_{ij}^{-*} = \sum_{(d,m) \in \mathcal{T}} f_{ij}^{-*}(d,m)$.

Meanwhile, by the flow conservation \eqref{flowconservation_f_domain}, we know that for all $j \in \mathcal{V}, (d,m) \in \mathcal{T}$,
\begin{align*}
    \sum_{i \in \mathcal{N}_i} t_i^{-*}(d,m) \phi_{ij}^{-*}(d,m) = t_j^{-*}(d,m) - r_j(d,m).
\end{align*}
Substitute above into the very last term in (\ref{proof_sum_1}) and cancel, we get
\begin{equation}
\begin{aligned}
    &\sum_{i\in\mathcal{V}} C^\prime_i(G_i) G_i^* + \sum_{(i,j) \in \mathcal{E}}D^\prime_{ij}(F_{ij})F^{-*}_{ij}
    \\  & \geq \sum_{i \in \mathcal{V}} \sum_{(d,m) \in \mathcal{T}} r_i(d,m)\frac{\partial T}{\partial r_i(d,m) } 
    \\ &- \sum_{i \in \mathcal{V}} \sum_{(d,m) \in \mathcal{T}} g_i^{*}(d,m) \frac{\partial T}{\partial t_i^+(d,m)}.
\end{aligned}
\label{proof_sum_1_final}
\end{equation}

Next, about the flow of computation result, multiply both side of (\ref{SuffCond_alt2}) by $\phi_{ij}^{+*}(d,m)$ and sum over $j \in \mathcal{N}_i $, we have
\begin{equation}
\begin{aligned}
    &\sum_{j \in \mathcal{N}_i}L_m^+D_{ij}^\prime(F_{ij}) \phi_{ij}^{+*}(d,m) 
    \\ \geq & \frac{\partial T}{\partial t_i^+(d,m)} - \sum_{j \in \mathcal{N}_i} \frac{\partial T}{\partial t_j^+(d,m)} \phi_{ij}^{+*}(d,m).
\end{aligned}    
\label{proof_ineqToEq2}
\end{equation}

Multiply both side by $t_i^{+*}(d,m) = \sum_{j \in \mathcal{N}_i}f_{ji}^{+*}(d,m) + g_i^{*}(d,m)$, sum over $(d,m) \in \mathcal{T}$ and $j \in \mathcal{V}$, we get
\begin{equation}
\begin{aligned}
    &\sum_{(i,j) \in \mathcal{E}}D_{ij}^\prime(F_{ij}) F_{ij}^{+*} 
    \\ &\geq \sum_{i \in \mathcal{V}} \sum_{(d,m) \in \mathcal{T}} t_i^{+*}(d,m) \frac{\partial T}{\partial t_i^+(d,m)} 
    \\ &- \sum_{(d,m)\in \mathcal{T}} \sum_{i \in \mathcal{V}} \sum_{j \in \mathcal{N}_i} t_i^{+*}(d,m) \frac{\partial T}{\partial t_j^+(d,m)} \phi_{ij}^{+*}(d,m) .
\end{aligned}    
 \label{proof_sum_2}
\end{equation}

By \eqref{flowconservation_f_domain}, we have for all $ j \in \mathcal{V}, (d,m) \in \mathcal{T}$,
\begin{align*}
    \sum_{i \in \mathcal{N}_i} t_i^{+*}(d,m) \phi_{ij}^{+*}(d,m) = t_j^{+*}(d,m) - g_j^{*}(d,m). 
\end{align*}

Substituting above into the very last term in (\ref{proof_sum_2}) and canceling, we get
\begin{align}
    \sum_{(i,j) \in \mathcal{E}} D_{ij}^\prime(F_{ij}) F_{ij}^{+*} \geq \sum_{i \in \mathcal{V}} \sum_{(d,m) \in S} g_i^{*}(d,m) \frac{\partial T}{\partial t_i^+(d,m)}.
     \label{proof_sum_2_final}
\end{align}

Summing up both side of (\ref{proof_sum_1_final}) and (\ref{proof_sum_2_final}), we have
\begin{equation}
\begin{aligned}
     &\sum_{i \in \mathcal{V}} C^\prime_i(G_i)G_i^* + \sum_{(i,j) \in \mathcal{E}} D_{ij}^\prime(F_{ij}) F_{ij}^{*} 
     \\ & \geq \sum_{i \in \mathcal{V}} \sum_{(d,m) \in \mathcal{T}} \frac{\partial T}{\partial r_i(d,m)} r_i(d,m).
\end{aligned}    
 \label{proof_ineq}
\end{equation}

Note that the equality would always hold in (\ref{proof_ineqToEq1}) and (\ref{proof_ineqToEq2}) if we substitute $\boldsymbol{\phi}^*$ with $\boldsymbol{\phi}$ in the above reasoning, as a consequence of (\ref{partial_D_r}) and (\ref{partial_D_t}).
Thus we have the following analogue of (\ref{proof_ineq}),
\begin{equation}
\begin{aligned}
    &\sum_{i \in \mathcal{V}} C^\prime_i(G_i)G_i + \sum_{(i,j) \in \mathcal{E}} D_{ij}^\prime(F_{ij}) F_{ij} 
    \\& = \sum_{i \in \mathcal{V}} \sum_{(d,m) \in \mathcal{T}} \frac{\partial T}{\partial r_i(d,m)} r_i(d,m).
\end{aligned}    
\label{proof_eq}
\end{equation}

Abstracting (\ref{proof_eq}) from (\ref{proof_ineq}), we show (\ref{proof_obj}) and complete the proof.

\subsection{Proof of Theorem \ref{thm:stability}}
\label{Proof:thm_stability}

We prove Theorem \ref{thm:stability} based on the following lemma.

\begin{lem}
\label{lem:stability}
Let $\boldsymbol{r} \in \mathcal{D}_{\boldsymbol{r}}$ and $\boldsymbol{\phi} \in \mathcal{D}_{\boldsymbol{\phi}}(\boldsymbol{r})$ satisfies condition \eqref{Condition_sufficient} given input $\boldsymbol{r}$, then there exist a function $\epsilon(\delta)$ for $\delta > 0$, 
such that $\lim_{\delta \to 0}\epsilon(\delta) = 0$, and the following holds:

For all $\Delta \boldsymbol{r}$ that $\left|\Delta \boldsymbol{r}\right| < \delta$ and $\left(\boldsymbol{r} + \Delta \boldsymbol{r}\right) \in \mathcal{D}_{\boldsymbol{r}}$, there exists a corresponding $\Delta\boldsymbol{\phi}$ that $\left|\Delta\boldsymbol{\phi}\right| < \epsilon(\delta)$, $\left(\boldsymbol{\phi} + \Delta\boldsymbol{\phi}\right) \in \mathcal{D}_{\boldsymbol{\phi}}\left(\boldsymbol{r} + \Delta \boldsymbol{r}\right)$, and $\left(\boldsymbol{\phi} + \Delta\boldsymbol{\phi}\right)$ satisfies condition \eqref{Condition_sufficient} given input $\left(\boldsymbol{r} + \Delta \boldsymbol{r}\right)$.
\end{lem}

Let \( \boldsymbol{r} \in \mathcal{D}_{\boldsymbol{r}} \) and \( \boldsymbol{\phi} \in F_{\text{suff}}(\boldsymbol{r}) \) be given. Let \( \{ \boldsymbol{r}^n \} \subset \mathcal{D}_{\boldsymbol{r}} \) be a sequence such that \( \boldsymbol{r}^n \to \boldsymbol{r} \). Define \( \delta_n = \|\boldsymbol{r}^n - \boldsymbol{r}\| \to 0 \).
By Lemma \ref{lem:stability}, for each \( n \), there exists \( \boldsymbol{\phi}^n \in F_{\text{suff}}(\boldsymbol{r}^n) \):
\[
\| \boldsymbol{\phi}^n - \boldsymbol{\phi} \| < \epsilon(\delta_n),
\]
and \( \lim_{n \to \infty} \epsilon(\delta_n) = 0 \). Hence, \( \boldsymbol{\phi}^n \to \boldsymbol{\phi} \).

Therefore, for every \( \boldsymbol{\phi} \in F_{\text{suff}}(\boldsymbol{r}) \) and every sequence \( \boldsymbol{r}^n \to \boldsymbol{r} \), we can construct a sequence \( \boldsymbol{\phi}^n \in F_{\text{suff}}(\boldsymbol{r}^n) \) such that \( \boldsymbol{\phi}^n \to \boldsymbol{\phi} \). This proves that \( F_{\text{suff}} \) is LHC at \( \boldsymbol{r} \).

\begin{proof}[proof of Lemma \ref{lem:stability}]
    
For analytical simplicity, we only prove for the case where $\Delta \boldsymbol{r}$ has only one non-zero element, that is, only $\Delta r_{i}(a,k) \neq 0$. General cases can be seen as a finite accumulations of this simple case.

Without loss of generality, we assume $|\mathcal{T}| = 1$ and assume a pure-routing scheme, i.e., the network only performs packet forwarding for only one task, without the need to conduct any computation. 
This simplifies the network formulation to Gallager's original setting \cite{gallager1977minimum}. 
Joint considering routing and computation is a naive extension of the pure routing case, as one can treat ``computing unit'' as one of network ``links'' .

Therefore, we omit the notation $(d,m)$ for simplicity.
We further assume all communication cost functions $D_{ij}(\cdot)$ are strictly convex.

Suppose $\boldsymbol{\phi}$ satisfies the sufficient condition \eqref{Condition_sufficient} with input rate $\boldsymbol{r}$, it is evident that $\boldsymbol{\phi}$ is loop-free.
When the input rate $r_i$ is increased by a sufficiently small $\Delta r_i$ ($\Delta r_i$ can be positive or negative, as long as the new input rate vector lies within stability region $\mathcal{D}_{\boldsymbol{r}}$. Without loss of generality, we assume $\Delta r_i > 0$), we apply Algorithm \ref{alg:sgp} one node at a time, starting from node $i$.
Consider the change of $[\delta_{ij}]$ for all $j$ when the forwarding strategy $\boldsymbol{\phi}$ is kept unchanged.
To break down the problem, consider the DAG (directed acyclic graph) constructed by $(i,j)\in\mathcal{E}$ such that $\phi_{ij} > 0$.

(1) If node $i$ is the destination node $d$, then $\delta_{ij}$ is not changed on any link.

(2) If node $i$ is one-hop away from $d$, then it must hold that $\phi_{id} = 1$ and $\phi_{ij} = 0$ for all other $j$. In this case, $\delta_{id} = D_{id}^\prime(f_{id})$, and $$\Delta \delta_{id} = D_{id}^\prime(f_{id} + \Delta r_i) - D_{id}^\prime(f_{id}) = \Delta r_i D^{\prime\prime}_{id}(f_{id}).$$
Therefore, we let $\delta^\prime_{id}$ be the marginal increase of $\delta_{id}$ due to the increase of $r_i$, and $$\delta^\prime_{id} = \frac{\Delta \delta_{id}}{\Delta r_i} = D^{\prime\prime}_{id}(f_{id}).$$
Moreover, let $\delta_i = \sum_{j} \phi_{ij} \delta_{ij}$, then $\delta^\prime_i = \sum_{j} \phi_{ij} \delta^\prime_{ij}$. In this case, $\delta_i^\prime = D^{\prime\prime}_{id}(f_{id})$.

(3) If node $i$ is more than one-hop away from $d$, then it holds that
\begin{equation}
\begin{aligned}
    &\delta_{ij}^\prime = D^{\prime\prime}_{ij}(f_{ij}) + \delta^\prime_j,
    \\& \delta_i^\prime =  \sum_{j} \phi_{ij} \delta^\prime_{ij}.
\end{aligned}
\label{proof_second_deri}
\end{equation}
We denote all paths from node $i$ to $d$ in the DAG by set $\mathcal{P}_i$, where each path $p \in \mathcal{P}_i$ is a sequence of nodes $(p_1,p_2,\cdots,p_{|p|})$ with $p_1 = i$, $p_{|p|} = d$ and $\phi_{p_k p_{k+1}} >0$ for $k = 1,\cdots,|p|-1$.
Therefore, by recursively applying \eqref{proof_second_deri} from node $d$ to node $i$ on the reverse direction for all path in $\mathcal{P}_i$, we have
\begin{equation}
\begin{aligned}
    \delta^\prime_i &= \sum_{p \in \mathcal{P}_i} \left(D^{\prime\prime}_{p_1p_2} + \left(D^{\prime\prime}_{p_2p_3} + \cdots \right)\phi_{p_2p_3}\right)\phi_{p_1p_2}
    \\ &= \sum_{p \in \mathcal{P}_i} \left(\sum_{k = 1}^{|p|-1} D^{\prime\prime}_{p_kp_{k+1}}\prod_{l = 1}^k \phi_{p_lp_{l+1}}\right)
\end{aligned}
\label{proof_delta_prime}
\end{equation}

Combining the above cases, we know that for any $i$, when $r_i$ in increased by a small amount $\Delta r_i$, the marginal cost $\delta_{ij}$ for an arbitrary $j$ is increased by
\begin{equation*}
    \Delta \delta_{ij} = \Delta r_i \left(D^{\prime\prime}_{ij} + \sum_{p \in \mathcal{P}_j} \left(\sum_{k = 1}^{|p|-1} D^{\prime\prime}_{p_kp_{k+1}}\prod_{l = 1}^k \phi_{p_lp_{l+1}}\right) \right).
\end{equation*}
Therefore, recall the algorithm update \eqref{variable_update}, and combined with the fact that $\boldsymbol{\phi}$ already satisfies \eqref{Condition_sufficient}. 
By the sensitivity of Lagrangian multipliers, the adjust amount $\Delta \phi_{ij} \equiv \phi^1_{ij} - \phi_{ij}$ is upper bounded by 
\begin{equation}
\label{proof_bound}
    \Delta \phi_{ij} \leq \alpha \Delta \delta_{ij}
\end{equation}
where the finite constant $\alpha$ is given by the Lagrangian multiplier of problem \eqref{variable_update}.
Moreover, it is shown by \cite{xi2008node} that Algorithm \ref{alg:sgp} converges linearly to the optimal solution satisfying \eqref{Condition_sufficient}.
By adopting section order methods, e.g., \cite{bertsekas1984second}, the rate of convergence can be enhanced to super-linear.
Therefore, let $\boldsymbol{\phi}^*$ be the convergent solution of Algorithm \ref{alg:sgp} after introducing $\Delta r_i$, there exist a finite scalar $M$ that
\begin{equation}
\label{proof_distance_bound}
    {|\boldsymbol{\phi}^{t+1} - \boldsymbol{\phi}^*|} \leq M {|\boldsymbol{\phi}^{t} - \boldsymbol{\phi}^*|},
\end{equation}
and there exists a scalar $\mu < 1$ such that 
\begin{equation}
\label{proof_convergence_rate}
    \lim_{t \to \infty} \frac{|\boldsymbol{\phi}^{t+1} - \boldsymbol{\phi}^*|}{|\boldsymbol{\phi}^{t} - \boldsymbol{\phi}^*|} = \mu.
\end{equation}

Combining \eqref{proof_bound}\eqref{proof_distance_bound}\eqref{proof_convergence_rate}, there exists a finite constant $C$ such that $| \boldsymbol{\phi}^* - \boldsymbol{\phi}| \leq C\Delta r_i$, i.e., $$\lim_{\Delta r_i \to 0}| \boldsymbol{\phi}^* - \boldsymbol{\phi}| = 0.$$

Therefore, there exist a function $\epsilon(\delta)$ for $\delta > 0$ continuous at $0$, 
and for $\Delta r_i$ that $\left|\Delta r_i\right| < \delta$, the corresponding new optimal solution $\boldsymbol{\phi}^*$ satisfies $| \boldsymbol{\phi}^* - \boldsymbol{\phi}| \leq \epsilon(\delta)$.
To generalize to multiple applications or multiple non-zero input rate $r_i$ changes, the analysis above still holds, as the constant $C$ would be the sum of all applications and all $\Delta r_i$, however, still finite.
To generalize to computation placement, one only needs to consider each computation step as a special link that goes back to the computation node itself, with a link cost associated.
\end{proof}


\bibliographystyle{IEEEtran}
\bibliography{References}


\end{document}